\documentclass[aps,prb,reprint,twocolumn,showpacs,superscriptaddress]{revtex4-1}
\usepackage{amsmath, amssymb, bm, amsthm}
\usepackage{graphicx}

\usepackage[usenames,dvipsnames]{color}			
\definecolor{purple}{rgb}{0.5,0,0.5}

\newcommand{\JJ}{\vec{\mathcal{J}}}
\newcommand{\QQ}{\vec{\mathcal{Q}}}
\newcommand{\LL}{\vec{\mathcal{L}}}
\newcommand{\MM}{\vec{\mathcal{M}}}
\newcommand{\FF}{\vec{\mathcal{F}}}
\newcommand{\GG}{\vec{\mathcal{G}}}

\def\br{{\bf r}}
\def\bR{{\bf R}}

\def\hall{\sigma_{xy}}
\def\Cnl{{\cal C}_\text{nonlocal}}

\begin{document}

\title{Lattice realization of a bosonic integer quantum Hall state - trivial insulator transition and relation to the self-dual line in the easy-plane NCCP1 model}
\author{Scott Geraedts}
\affiliation{Department of Physics, Princeton University, Princeton, New Jersey 08544, USA}
\affiliation{Department of Electrical Engineering, Princeton University, Princeton, New Jersey 08544, USA}
\author{Olexei I.~Motrunich}
\affiliation{Department of Physics, California Institute of Technology, Pasadena, California 91125, USA}
\affiliation{Institute for Quantum Information and Matter, Caltech, Pasadena, California 91125, USA}

\date{\today}

\begin{abstract}
We provide an explicit lattice model of bosons with two separately conserved boson species [$U(1)\times U(1)$ global symmetry] realizing a direct transition between an integer quantum Hall effect of bosons and a trivial phase, where any intermediate phase is avoided by an additional symmetry interchanging the two species.
If the latter symmetry is absent, we find intermediate superfluid phases where one or the other boson species condenses.
We know the precise location of the transition since at this point our model has an exact nonlocal antiunitary particle-hole-like symmetry that resembles particle-hole symmetry in the lowest Landau level of electrons.
We exactly map the direct transition to our earlier study of the self-dual line of the easy-plane NCCP1 model, in the mathematically equivalent reformulation in terms of two (new) particles with $\pi$ statistics and identical energetics.
While the transition in our model is first order, we hope that our mappings and recent renewed interest in such self-dual models will stimulate more searches for models with a continuous transition.
\end{abstract}

\maketitle

\section{Introduction}

For much of the history of condensed matter physics, different phases of matter were understood as being related to breaking different symmetries.  
In more recent times, ``topologically ordered'' phases outside of this paradigm, such as the fractional quantum Hall effect, have been discovered.\cite{PrangeGirvin, DasSarmaBook, Wen_book}
Even more recently, the condensed matter community has realized that there is another type, the ``symmetry protected topological phases'' (SPT) which,\cite{Chen2012_Science, Chen2013_PRB, Senthil_SPTreview2015} though not topologically ordered in the sense of the fractional quantum Hall effect, are distinct in the presence of some symmetry from the ``topologically trivial'' phase with that symmetry.

Much is understood about the phase transitions between different conventional phases (i.e., those related by symmetry breaking), but phase transitions between SPT phases is still largely uncharted territory.  There are two types of such transitions.  In the simplest case, the symmetry protecting the SPT phases is broken, and there is a transition to a topologically trivial phase with less symmetry.  The theory of such a transition can be studied in a variety of models\cite{Gu2009, Jiang2010, Lange2015} and seems to possess the same properties as a transition between topologically trivial phases where the same symmetry is broken.  The more challenging case is the transition between the SPT phase and the topologically trivial phase, where {\it no symmetry is broken}.  

One of the topological phases thought to exhibit such a transition is the bosonic integer quantum Hall effect (BIQHE).\cite{LuVishwanath2012, SenthilLevin2012, FQHE}
The BIQHE has been realized numerically in both continuum \cite{Furukawa-PhysRevLett.111.090401, Wu-PhysRevB.87.245123, Regnault-PhysRevB.88.161106, Geraedts2017} and lattice \cite{Moller:2009p184, FQHE, Sterdyniak-PhysRevLett.115.116802, He2015, Zeng-PhysRevB.93.195121} models, but generically these are not expected to realize a direct transition between the BIQHE and the trivial insulator with the same symmetry.\cite{GroverVishwanath2013, LuLee2014_QPT}
Recent theoretical and numerical studies\cite{SlagleYouXu2015, HeWuYouXuMengLu2016, BiZhangYouYoungBalentsLiuXu2017} of bilayer graphene as a platform for bosonic SPT states suggested a second-order bosonic topological-trivial transition, although the accessed sizes are perhaps too small to determine critical properties.
In other works, DMRG and coupled-wire studies of a hard-core boson model with correlated hopping on a honeycomb lattice\cite{He2015, FujiHe2016} found a direct transition between BIQHE states with $\hall = 2$ and $\hall = -2$, but more detailed studies are needed to establish the nature of the transition.

In this paper we show how a model introduced in Ref.~\onlinecite{FQHE} can realize a direct transition between the BIQHE and the trivial insulator with the same symmetry.
Our model has two species of separately conserved bosons with short-range interactions that break conventional time-reversal symmetry but preserve an alternate antiunitary symmetry that allows a sign-free reformulation and Monte Carlo studies already employed in previous works.\cite{Loopy, short_range3, FQHE, jongyeon}
Both species are at integer filling due to an additional local unitary particle-hole symmetry, which however is not an obstacle to producing BIQHE or its fractionalized cousins, i.e., ``symmetry-enriched topological phases'' (SET).
We then show that requiring an interchange symmetry between the two boson species is crucial for realizing the direct transition between the BIQHE and trivial phase.
Furthermore, we can impose an interesting nonlocal antiunitary particle-hole-like symmetry that puts our model \emph{exactly} at the transition.
Such a model with the above symmetries placed exactly at the transition is in fact equivalent to so-called easy-plane NCCP1 model (EP-NCCP1) at exact self-duality,\cite{shortlight, deccp_science, deccp_prb} and we use this mapping to connect our model to a previous numerical study\cite{Loopy} of the self-dual line in the EP-NCCP1 model.
While this earlier study found that the transition in the specific model is first order, the detailed understanding in the present paper allows us to propose more general models that can be similarly placed exactly at the transition and may realize continuous transitions.

It is useful to relate our paper to recent advances in the understanding of interplay of symmetries and dualities.\cite{Son, WangSenthil2015, MetlitskiVishwanath2015, Mross2016_diracduality, SeibergSenthilWangWitten2016, KarchTong2016, WangNahum2017}
Thus, the nonlocal antiunitary particle-hole-like symmetry mentioned above is a bosonic analog of the electronic particle-hole symmetry in the lowest Landau level, which has recently attracted much interest.\cite{Son, WangSenthil2015, MetlitskiVishwanath2015, Geraedts2016_science}
The bosonic analog introduced in Ref.~\onlinecite{WangSenthil2016_CFLs} interchanges the BIQHE and trivial insulator phases, and from this perspective it is natural that this symmetry places our model exactly at the transition between the two phases.
In the conventional fractional quantum Hall setting for bosons, exact realization of the particle-hole symmetry has been elusive, while it is quite natural in our model.
Next, in our model the species-interchange symmetry of bosons corresponds to exact self-duality in the EP-NCCP1 model, i.e., a symmetry in one set of variables maps to a duality transformation in another set.
Explorations of such an interplay is also a very active topic, and our model provides an interesting exact example that can also be of practical use for numerical studies of the transition in the EP-NCCP1 model (which we already employed in Ref.~\onlinecite{Loopy}).
Finally, the self-dual EP-NCCP1 model has been related to self-dual QED$_3$ with two species of Dirac fermions,\cite{Senthil2006_theta, WangNahum2017, Mross2017} so finding a continuous such transition would have direct implications for possible critical field theory of such QED$_3$.

The paper is presented as follows:
In Sec.~\ref{sec:model} we review how the model is constructed, while in Sec.~\ref{sec:symmetry} we discuss the model's symmetries.
In Sec.~\ref{sec:phase} we discuss the phase diagram of our model.
We support this phase diagram by summarizing a number of previous Monte Carlo studies.\cite{Loopy, short_range3, FQHE, jongyeon}
Finally in Sec.~\ref{sec:relations} we explicitly show how the model is connected to the EP-NCCP1 model, and also how it connects to the model studied numerically in Ref.~\onlinecite{Loopy}; while in Sec.~\ref{sec:discussion} we conclude with some discussion and outlook.
In Appendix~\ref{app:H2topophases} we illustrate how we can argue for bosonic SPT and SET phases directly in our Hamiltonian.

\section{Model Hamiltonian}
\label{sec:model}

\begin{figure}
\includegraphics{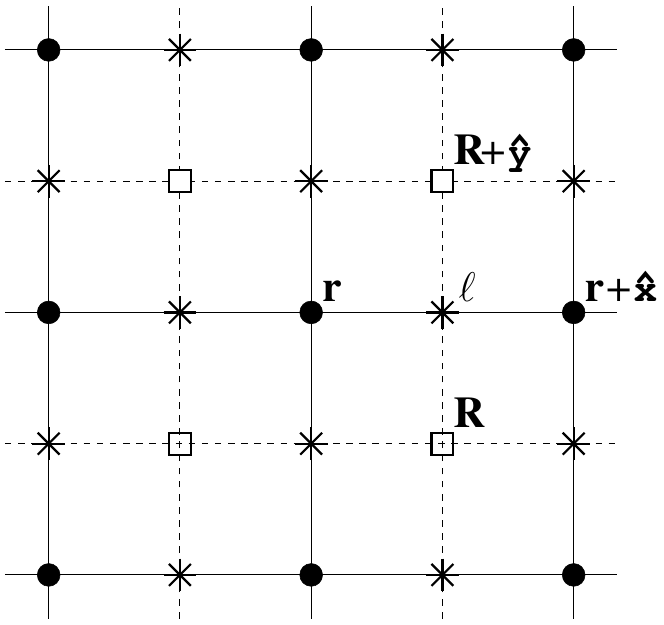}
\caption{The lattice on which the Hamiltonian in Eq.~(\ref{H}) is defined.  There are two interpenetrating cubic lattices with sites labeled by the coordinates $\br$ and $\bR$.  On the sites of each lattice there are $U(1)$ quantum rotors.  On the intersection points of the links of the two lattices there are harmonic oscillators, whose ``position'' and ``momentum'' variables are conveniently thought of as fields residing on the oriented links of the direct and dual lattices respectively.
}
\label{fig:ham}
\end{figure}

We now describe in detail the model used in this paper, which was originally introduced in Ref.~\onlinecite{FQHE}.  The model is defined on two interpenetrating square lattices, as seen in Fig.~\ref{fig:ham}.  On each site of one lattice we place a $U(1)$ quantum rotor, which can be described by a compact phase variable $\hat{\phi}_1(\br)$, where $\br$ indicates a site of one of the lattices (from now on called the ``direct lattice'').  The rotors can also be described by conjugate number variables $\hat{n}_1(\br)$ which are integer-valued and satisfy $[\hat{\phi}_1(\br), \hat{n}_1(\br')] = i \delta_{\br \br'}$.  On the other lattice we put a different species of rotors, defined by the variables $\hat{\phi}_2(\bR)$, $\hat{n}_2(\bR)$, where $\bR$ indicates a site on the other lattice (from now on called the ``dual lattice'').  We can also think of these rotors as representing two species of bosons, where $\hat n$ is the number of bosons on each site and $\hat\phi$ is the phase of those bosons.

As can be seen in Fig.~\ref{fig:ham}, the links of the two lattices intersect, and at each such intersection point we place a harmonic oscillator, described by conjugate real-valued variables $\chi_\ell$, $\pi_\ell$, where $\ell$ labels a link of either lattice, and 
\begin{equation}
[\hat\chi_\ell, \hat\pi_{\ell^\prime}] = i \delta_{\ell, \ell^\prime} ~.
\end{equation}
A given link can be described by a combination of a position and direction on either lattice, i.e., we can replace $\ell$ by $\br, j$ ($j$ is a direction, $\ell = \langle \br, \br + \hat{j} \rangle$) or $\bR, j^\prime$ ($j^\prime$ is a direction perpendicular to $j$; $\ell = \langle \bR, \bR + \hat{j}^\prime \rangle$).  We can then define variables associated with oriented links on the direct and dual lattices,
\begin{equation}
\hat{\alpha}_{1j}(\br) = \hat{\chi}_\ell ~, ~~ \hat{\alpha}_{2j^\prime}(\bR) = 
\left\{
\begin{array}{c}
\hat{\pi}_\ell ~~{\rm if}~ \hat{j}^\prime = \hat{x} \\
-\hat{\pi}_\ell ~~{\rm if}~ \hat{j}^\prime = \hat{y} 
\end{array}
\right. .
\end{equation}
The commutation relations for the $\hat{\alpha}$-s are therefore:
\begin{equation}
[\hat{\alpha}_{1j}(\br), \hat{\alpha}_{2j^\prime}(\bR)] = -i \epsilon_{jj^\prime} \delta_{\br + \hat{j}/2 = \bR + \hat{j}^\prime/2} ~,
\label{alphacomm}
\end{equation}
where $\epsilon_{jj^\prime}$ is the antisymmetric tensor.

With all our degrees of freedom defined, we can now write the Hamiltonian:
\begin{eqnarray}
\label{H}
\hat{H} &=& \hat{H}_{h1} + \hat{H}_{h2} + \hat{H}_{u1} + \hat{H}_{u2} + \hat{H}_\alpha ~, \\
\hat{H}_{h1} &=& -\sum_{\br, j} h_1 \cos\left[\nabla_j \hat{\phi}_1(\br) - \sqrt{\frac{2\pi}{\eta}} \hat{\alpha}_{1j}(\br) \right] ~, \\
\hat{H}_{h2} &=& -\sum_{\bR, j} h_2 \cos\left[\nabla_j \hat{\phi}_2(\bR) - \sqrt{\frac{2\pi}{\eta}} \hat{\alpha}_{2j}(\bR)\right] ~, \\
\hat{H}_{u1} &=& \frac{1}{2} \sum_\br u_1 \left[\hat{n}_1(\br) + \sqrt{\frac{\eta}{2\pi}} ({\bm \nabla} \wedge \hat{\bm \alpha}_2)(\br) \right]^2 ~, \\
\hat{H}_{u2} &=& \frac{1}{2} \sum_\bR u_2 \left[\hat{n}_2(\bR) + \sqrt{\frac{\eta}{2\pi}} ({\bm \nabla} \wedge \hat{\bm \alpha}_1)(\bR) \right]^2 ~, \\
\hat{H}_\alpha &=& \sum_\ell \left[ \frac{\kappa_1}{2} \hat{\alpha}_1(\ell)^2 + \frac{\kappa_2}{2} \hat{\alpha}_2(\ell)^2 \right] ~.
\end{eqnarray}
Here ${\bm \nabla} \wedge \hat{\bm \alpha} \equiv \nabla_x \hat\alpha_y - \nabla_y \hat\alpha_x$ denotes the spatial curl, which is naturally centered at dual or direct lattice sites for ${\bm \hat\alpha}_1$ or ${\bm \hat\alpha}_2$, respectively.  
The parameter $\eta$ in this model eventually determines which topological phases will be realized.
To understand the physics of this Hamiltonian, we start with the terms $H_{h1}$, $H_{h2}$.  What these terms do is tie the curls ${\bm \nabla} \wedge \hat{\bm \alpha}_{1,2}$ to the vorticities of the phase variables $\hat{\phi}_{1,2}$.  Then, terms $H_{u2}$, $H_{u1}$ force the number variables of the opposite species, $\hat n_{2,1}$, to be proportional to these curls.  The net result is that the Hamiltonian tries to tie $\eta$ bosons of one species to a vortex of the other species.
When $\eta = 1$ this therefore implements boson-vortex binding which gives rise to the bosonic integer quantum Hall effect (BIQHE).\cite{LuVishwanath2012, SenthilLevin2012, FQHE}
When $\eta$ is a rational number, such a Hamiltonian produces bound states of multiple bosons and vortices (e.g., $c$ bosons and $d$ vortices for $\eta = c/d$ with mutually prime integers $c$ and $d$); this leads to a wide variety of fractionalized topological phases, which are the symmetry-enriched topological phase versions of the BIQHE.\cite{FQHE}

The Hamiltonian in Eq.~(\ref{H}) cannot be solved exactly (but see Appendix~\ref{app:H2topophases} for analysis in special limit).
In Ref.~\onlinecite{FQHE} we developed its imaginary-time path integral, which, for judiciously chosen parameters, led to the following action in (2+1) dimensions:
\begin{eqnarray}
S_{JJ} &=& \frac{1}{2} \sum_{r,r'} v_1(r-r')~ \JJ_1(r) \cdot \JJ_1(r') \label{action} \\
&& + ~ \frac{1}{2} \sum_{R,R'} v_2(R-R')~ \JJ_2(R) \cdot \JJ_2(R') \nonumber \\
&& + ~ i \sum_{R,R'} w(R-R')~ [\vec{\nabla} \times \JJ_1](R) \cdot \JJ_2(R') ~. \nonumber
\end{eqnarray}
The coordinates $r$ in the above equation now represent a (2+1)-dimensional space-time lattice made up of both the spatial position $\br$ as well as imaginary time $\tau$, and similarly for the coordinates $R$.
As discussed in Ref.~\onlinecite{FQHE}, the imaginary time position on the $R$ lattice is naturally displaced by $1/2$ relative to the $r$ lattice, so we obtain two (2+1)-dimensional lattices which are dual to each other.

The variables $\JJ_{1,2}$ are conserved space-time currents, $\vec{\nabla} \cdot \JJ_{1,2} = 0$, and represent the worldlines of the bosons $\hat{n}_{1,2}$ in Eq.~(\ref{H}), and they interact with each other through the potentials $v_{1,2}$ (for intraspecies interactions) and $w$ (for interspecies interactions).
These potentials have the following form in momentum space:
\begin{eqnarray}
&& v_{1/2}(k) = \frac{\lambda_{2/1}}{\lambda_1 \lambda_2 + \frac{\eta^2 |\vec{f}_k|^2}{(2\pi)^2}} ~, \label{vintro} \\
&& w(k) = \frac{-\eta}{2\pi} \frac{1}{\lambda_1 \lambda_2 + \frac{\eta^2 |\vec{f}_k|^2}{(2\pi)^2}} ~. \label{tintro}
\end{eqnarray}
The parameters $\lambda_{1/2}$ can be expressed in terms of the parameters in Eq.~(\ref{H}) as well as the imaginary time discretization $\delta\tau$.
For judiciously chosen parameters, we can make the system isotropic in space and imaginary time: 
\begin{equation}
\lambda_1 = \delta\tau \kappa_1 \frac{\eta}{2\pi} = \frac{1}{\delta\tau u_1} ~,
\end{equation}
and similarly for $\lambda_2$.
We also defined $f_{k,\mu} \equiv 1 - e^{i k_\mu}$, so $i \vec{f}_k$ is the lattice version of the momentum $\vec{k}$; thus, $|\vec{f}_k|^2 = \sum_\mu (2 - 2\cos k_\mu)$.
The parameters $h_1$, $h_2$ in Eq.~(\ref{H}) are assumed to be very large.

In Ref.~\onlinecite{FQHE} we showed how actions such as Eq.~(\ref{action}) can be studied in sign-free Monte Carlo.
The numerical results of the present paper came from carrying out such simulations.
For both the Monte Carlo simulations and analytical progress, it is very useful to perform a duality transformation which is an exact rewriting of the partition sum in terms of new (``dual'' or ``vortex'') variables as defined in Appendix A of Ref.~\onlinecite{FQHE} (which also gives precise conditions on the currents for the transformation to be exact). 
By dualizing one species, $\JJ_1 \to \QQ_1$, while leaving the second species untouched, our model Eq.~(\ref{action}) with potentials Eqs.~(\ref{vintro}) and (\ref{tintro}) leads to action\cite{FQHE}
\begin{eqnarray}
S_{QJ} = \sum_k \frac{(2\pi)^2 \lambda_1}{2 |\vec{f}_k|^2} |\QQ_1(k)|^2
+ \sum_{R} \frac{|\JJ_2(R) - \eta \QQ_1(R)|^2}{2\lambda_2}.~~~~ \label{QJreal}
\end{eqnarray}
If $\eta$ is a rational number, $\eta = c/d$, then for small $\lambda_{1,2}$ this action binds $d$ vortices to $c$ bosons, and proliferates the resulting objects.

\section{Symmetries}
\label{sec:symmetry}
The above model has a number of symmetries which are vital to understanding it.  Since both species of bosons are conserved, there are two $U(1)$ symmetries.  In the loop model of Eq.~(\ref{action}), these symmetries force the $\JJ$ variables to form closed loops ($\vec{\nabla} \cdot \JJ = 0$), while in the Hamiltonian of Eq.~(\ref{H}) they appear as the invariance under $\hat{\phi}_1(\br) \to \hat{\phi}_1(\br) + \gamma_1$ (with position-independent $\gamma_1$) and similarly for $\hat{\phi}_2(\bR)$. 

Our model also has a unitary particle-hole symmetry ${\cal C}_\text{unitary}$, defined as
\begin{eqnarray}
{\cal C}_\text{unitary} ~:~ \JJ_s & ~\to~ & -\JJ_s ~, \label{unitaryPH} \\
\hat{n}_s & ~\to~ & -\hat{n}_s ~, \quad \hat{\phi}_s ~\to~ -\hat{\phi}_s ~, \\
\hat{\bm \alpha}_s & ~\to~ & -\hat{\bm \alpha}_s ~,
\end{eqnarray}
for both species $s = 1, 2$, where the first line refers to invariance of the Euclidean action while the second line specifies the symmetry in the Hamiltonian language.

Additionally, our model has an antiunitary symmetry ${\cal T}_{-+}$ which acts as particle-hole on one species but does not change the particle number of the other species:
\begin{eqnarray}
{\cal T}_{-+} ~:~ \JJ_1 & ~\to~ & -\JJ_1 ~, \label{antiunitarySym} \\
\JJ_2 & ~\to~ & \JJ_2 ~, \\
i & ~\to~ & -i ~, \\
\hat{n}_1 & ~\to~ & -\hat{n}_1 ~, \quad \hat{\phi}_1 ~\to~ \hat{\phi}_1 ~, \\
\hat{n}_2 & ~\to~ & \hat{n}_2 ~, \quad \hat{\phi}_2 ~\to~ -\hat{\phi}_2 ~ ,\\
\hat{\bm \alpha}_1 & ~\to~ & \hat{\bm \alpha}_1 ~, \quad \hat{\bm \alpha}_2 ~\to~ -\hat{\bm \alpha}_2 ~.
\end{eqnarray}
This symmetry is the reason we are able to simulate the model in a sign-free way:
It takes $\QQ_1 \to \QQ_1$, and since the symmetry is antiunitary this implies that the action $S_{QJ}$ in terms of $\QQ_1$ and $\JJ_2$ variables is real-valued and therefore simulable in Monte Carlo. We can see this explicitly for the action Eq.~(\ref{QJreal}) obtained for the model potentials in Eqs.~(\ref{vintro}) and (\ref{tintro}), but this holds more generally as long as we have ${\cal T}_{-+}$.

The above symmetries are the only independent symmetries enjoyed by all of our models.
Note that we do not have ``conventional time reversal'' symmetry that would act identically on both species (such as $\JJ_s \to \JJ_s, i \to -i$), which is why our models can realize integer and fractional quantum Hall phases with nonzero $\sigma_{xy}$ (more precisely, nonzero cross-species transverse response $\sigma_{xy}^{12}$, see Ref.~\onlinecite{FQHE} for details).

The loop model of Eq.~(\ref{QJreal}) has another interesting property in that the action remains unchanged under the following transformation of variables and parameter $\eta$:
\begin{eqnarray}
{\cal C}_\text{nonlocal} ~:~ \JJ_2 & ~\to~ & \QQ_1 - \JJ_2 ~, \label{nonlocalPH} \\
\QQ_1 & ~\to~ & \QQ_1 ~, \\
\eta & ~\to~ & 1 - \eta ~.
\end{eqnarray}
Since $\eta = 1/2$ maps to itself, this transformation can be viewed as a symmetry of the model with $\eta = 1/2$.
Unlike the symmetries discussed above, this transformation is nonlocal in the $\JJ_1$ and $\JJ_2$ variables, and therefore it is difficult to see how it is realized microscopically in the Hamiltonian Eq.~(\ref{H}) (but see the next subsection).
Here this property implies that as we vary $\eta$ in the loop model, we expect a transition exactly at $\eta = 1/2$, which we will argue in the next section is between a trivial insulator ($\sigma_{xy} = 0$) and a BIQHE insulator ($\sigma_{xy} = 2$).

Running somewhat ahead, we note that this property has features expected of a bosonic analog of the particle-hole transformation of electrons in the lowest Landau level:\cite{WangSenthil2015}
As we will show below, the above transformation interchanges the trivial insulator and BIQHE insulator; more generally, it maps a fractional quantum Hall state of bosons with $\sigma_{xy} = 2c/d$ to a new state with $\sigma_{xy} = 2 - 2c/d$.
The change in sign on $\JJ_2$ in Eq.~(\ref{nonlocalPH}) signifies the particle-hole aspect of the symmetry for the second species.
Furthermore, we expect this property to have identical manifestation when expressed in $\JJ_1$ and $\QQ_2$ variables.
That is, there is a change in sign for both boson currents and no change in sign for both vortex currents, and this implies that the transformation is antiunitary, which we will see explicitly in the Hamiltonian language at $\eta = 1/2$.

Finally, our model has an additional ``species interchange symmetry'' when the parameters satisfy $\lambda_1 = \lambda_2$ (or equivalently $h_1 = h_2$, $u_1 = u_2$, $\kappa_1 = \kappa_2$):
\begin{eqnarray}
{\cal R} ~:~ \JJ_1 & ~ \leftrightarrow ~ & \JJ_2 ~, \label{interchange} \\
\hat{n}_1 & ~ \leftrightarrow ~ & \hat{n}_2 ~, \quad \hat{\phi}_1 ~ \leftrightarrow ~ \hat{\phi}_2 ~, \\
\hat{\bm \alpha}_1 & ~\leftrightarrow ~ & \hat{\bm \alpha}_2 ~.
\end{eqnarray}
Since the variables of different species live on different lattices, this symmetry implicitly also involves a translation by half a unit cell in all directions.

\subsection{Hamiltonian formulation of the nonlocal particle-hole symmetry at $\eta = 1/2$}
The BIQHE can be understood\cite{FQHE} through Eq.~(\ref{QJreal}) as a condensate of bound states of bosons $\JJ_2$ and vortices $\QQ_1$.
Thinking in this context, the difference $\QQ_1 - \JJ_2$ is nonzero when a vortex does not have a bound boson and is reminiscent of a ``hole'' in a boson IQHE state.
The transformation ${\cal C}_\text{nonlocal}$ therefore seems like a bosonic equivalent to the fermionic particle-hole symmetry in a Landau level.
The latter symmetry, though nonlocal, can be understood by restricting to a single Landau level, and it is natural to ask if the same phenomenon is possible in our model.

At $\eta = 1/2$, we can restrict to a Landau-level-like structure by taking $h_1, h_2 \to \infty$.
This limit is tractable since $\hat{H}_{h1}$ and $\hat{H}_{h2}$ commute when $\eta = 1/2$.
These terms lock respective ${\bm \alpha}_s$ to ${\bm \nabla} \phi_s$ and therefore effectively lock ${\bm \nabla} \wedge {\bm \alpha}_s$ to the vorticity of $\phi_s$.
Indeed, we have, e.g., $\sqrt{2\pi/\eta} \, \alpha_{1j}(\br) = \nabla_j \phi_1(\br) - 2\pi p_{1j}(\br)$ with integer-valued $p_{1j}(\br)$, and hence 
\begin{equation}
\sqrt{2\pi/\eta} \, {\bm \nabla} \wedge {\bm \alpha}_1 = -2\pi {\bm \nabla} \wedge {\bm p}_1 \equiv -2\pi Q_1 ~,
\end{equation} 
for $Q_1$ some integer.
If we require $\alpha_{1j}(\br)$ to be as small as possible, which is what the energetics term $H_\alpha$ wants, then $Q_1$ coincides with the commonly used definition of vorticity of $\phi_1$.
Of course, the energetics requirement on the smallness of $\alpha_1$ is only ``soft'', but the precise interpretation of the integers $Q_1$ (and similarly arising integers $Q_2$) is not important in what follows.

Thus, the large $h$ limit allows us to consider a restricted Hilbert space defined by $\exp(i \sqrt{2\pi/\eta} \, {\bm \nabla} \wedge {\bm \alpha}_s) = 1$ on each direct or dual lattice placket corresponding to fluxes of ${\bm \alpha}_1$ or ${\bm \alpha}_2$ respectively.
With this restriction made, we can rewrite the $\hat H_{u1}$ term in Eq.~(\ref{H}) as:
\begin{equation}
\hat{H}_{u1} = \frac{1}{2} \sum_\br u_1 \left[\hat{n}_1(\br) - \eta Q_{2}(\br) \right]^2,
\end{equation}
and similarly for $\hat{H}_{u2}$ (simply interchanging species labels $1$ and $2$).
When $\eta = 1/2$, our model now has a symmetry $n_s \to Q_{\bar{s}} - n_s$, similar to the symmetry $\JJ_s \to \QQ_{\bar{s}} - \JJ_s$ of the loop model, where $\bar{s}$ means the other species relative to $s$.
This symmetry needs to be antiunitary since we do not want it to transform ${\bm \alpha}_s$ and hence $\phi_s$.
This is then a precise realization of antiunitary particle-hole transformation in our Hamiltonian, and the fact that we can define it only upon some restriction on the Hilbert space is reminiscent of the status of the electronic particle-hole symmetry in a Landau level.

We remark that in terms of the loop variables, the nonlocal particle-hole transformation applies for any $\eta$, but in the Hamiltonian model the symmetry only holds for $\eta = 1/2$.
This is because the above projection obtained by taking $h_1, h_2 \to \infty$ is a different projection for different $\eta$.
Hence, the restricted Hilbert spaces are different for different $\eta$ (including also cases $\eta = \eta_0$ and $\eta = 1 - \eta_0$), and we do not have a unique restricted Hilbert space (same for all $\eta$) in which we could define a particle-hole transformation as above.
In this way, our model is different from the traditional quantum Hall problem.
However, if we confine ourselves to $\eta = 1/2$ where we do have exact nonlocal particle-hole symmetry, conceptually, this difference is not an issue. 

Appendix~\ref{app:H2topophases} contains an additional example working in the Hamiltonian language, where we analyze fractionalized insulator at $\eta = 1/d$.
In the remainder of the main text, we will use the loop models to explore the transition between the trivial and integer quantum Hall effect and its relation to other interesting theories.

\section{Phase Diagram}
\label{sec:phase}

\begin{figure*}
\includegraphics[width=0.9\linewidth]{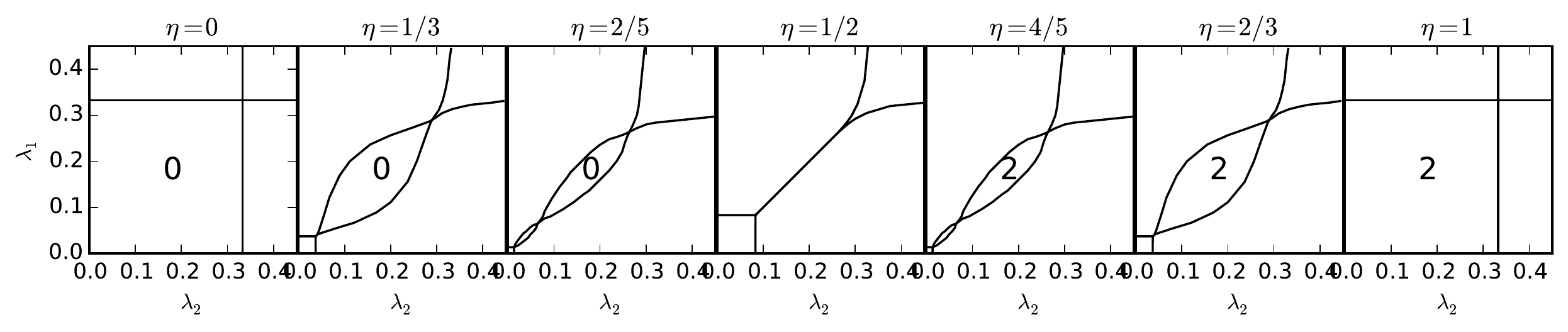}
\caption{Phase diagrams for the model in Eqs.~(\ref{action})-(\ref{tintro}),\cite{Loopy, short_range3, jongyeon} as a function of $\lambda_1$, $\lambda_2$ and for a variety of different $\eta$.
A description of each of the phases is given in the main text.
The phases labeled with numbers are either the trivial insulator with Hall conductivity $0$ or the BIQHE with Hall conductivity $2$.}
\label{fig:fixedeta}
\end{figure*}

We now review the phase diagram of the loop model in Eq.~(\ref{action}) as a function of the parameters $\lambda_1$, $\lambda_2$, and $\eta$.
Figure~\ref{fig:fixedeta} shows a number of cuts through this parameter space, each of which has fixed $\eta$ and varies $\lambda_1$ and $\lambda_2$.

\subsection{Features common to all values of $\eta$} 
There are some phases which are the same in all of the panels of Fig.~\ref{fig:fixedeta}.
In particular, when $\lambda_1, \lambda_2$ are comparable and very large, we can see from Eqs.~(\ref{vintro}) and (\ref{tintro}) that the interactions all become very small.
This leads to a proliferation of $\JJ_1$ and $\JJ_2$ worldlines, and therefore in this case the system is a superfluid breaking both $U(1)$ symmetries, regardless of $\eta$.

On the other hand, when, say, $\lambda_1$ is very large but $\lambda_2$ is not, then the $\JJ_1$ variables see a very small interaction while the $\JJ_2$ variables see a large interaction.
Therefore the first species is superfluid while the second is a trivial insulator, i.e., only one $U(1)$ symmetry is broken.

\subsection{$\eta = 0$}
In Fig.~\ref{fig:fixedeta} we can see that for $\eta = 0$ there are four phases. 
Three of them are the superfluids described above, so the only case left to explain is when $\lambda_1$ and $\lambda_2$ are both small.
From Eqs.~(\ref{vintro}) and (\ref{tintro}) we can see that when this happens and $\eta = 0$, both species of bosons see a large interaction potential, and there will be no worldlines of either species.
Therefore this phase is a trivial insulator of both species, and no symmetries are broken.
Equivalently, in this phase the worldlines of vortices, $\QQ_{1,2}$, have proliferated [which can be also established from Eq.~(\ref{QJreal}) where $\QQ_1$ see small potential while $\JJ_2$ see large potential].

\subsection{$\eta = 1$}
The property $\Cnl$ forces the phase diagram at $\eta = 1$ to have transitions in exactly the same places as that at $\eta = 0$.
However, the nature of the phases can be different.
In particular, when $\lambda_1$, $\lambda_2$ are both small we can see from Eq.~(\ref{QJreal}) that objects with $\QQ_1 = \JJ_2$ will have only a small energy cost, leading to a proliferation of bound states of bosons and vortices.
This leads to the boson integer quantum Hall effect (BIQHE) with $\hall = 2$, see Ref.~\onlinecite{FQHE} for details. 

\subsection{$0 < \eta < 1/2$}
\label{subsec:fract_eta}
When $\eta$ takes a fractional value---for concreteness, let $\eta = c/d$ with $c$ and $d$ mutually prime integers,---we can again understand the phase at small $\lambda_1$ and $\lambda_2$ using Eq.~(\ref{QJreal}), but this time there is a binding of $d$ vortices to $c$ bosons.
This leads to a topologically ordered generalization of the BIQHE phase.
The phase has $\hall = 2\eta$, gapless counterpropagating edge modes, and hosts quasiparticles which have fractional charge $1/d$ and fractional mutual statistics $2\pi b/d$ between different species (where integers $a, b$ are defined from the mutual primeness of $c$ and $d$ via $ad - bc = 1$), see Ref.~\onlinecite{FQHE} for details.

When $\eta$ has the form $1/d$, such as $\eta = 1/3$ in Fig.~\ref{fig:fixedeta}, in addition to the topologically ordered phase mentioned above (and the various superfluids at large $\lambda$), there is another phase at intermediate $\lambda_1,\lambda_2$.
This phase exists because for such $\eta$ to get the topological phase we need to proliferate composites containing $d$ vortices $\QQ_1$. 
However, the first term of Eq.~(\ref{QJreal}) tries to prevent this condensation.
If $\lambda_c$ is the value of $\lambda_1$ at which the trivial phase appears at $\eta = 0$, then condensing a $d$-fold composite of $Q_1$ will require $\lambda_1 \lesssim \lambda_c/d^2$.
However, above this value and below $\lambda_c$ (i.e., in the regime $\lambda_c/d^2 \lesssim \lambda_1 \sim \lambda_2 \lesssim \lambda_c$), single $\QQ_1$ variables can condense, while $\JJ_2$ variables can stay gapped. 
The resulting phase is therefore the same trivial insulator that occurred at small $\lambda_{1,2}$ and $\eta = 0$.

For other rational $\eta = c/d$, such as $\eta = 2/5$ shown in Fig.~\ref{fig:fixedeta}, the phase diagram has a more complicated structure.
We still have the topologically ordered phase at small $\lambda_{1,2} \lesssim \lambda_c/d^2$ with $\hall = 2 \eta$ described above.
However, near the line $\lambda_1 = \lambda_2$, and between $\lambda_c$ (above which we have the superfluid of both species) and $\lambda_c/d^2$ (below which we have the topologically ordered phase), we now have additional phases.
These phases can be understood as condensing more and more complicated bound states of $\QQ_1$ and $\JJ_2$ variables.
Just below $\lambda_c$, only single-strength $\QQ_1$ vortices can condense, leading to the trivial insulator. 
At smaller $\lambda$ we can condense composites with multiple $\QQ_1$; e.g., for $\eta = 2/5$ in the region between (approximately) $\lambda_c/25$ and $\lambda_c/4$ we can condense objects with $\QQ_1 = 2, \JJ_2 = 1$, leading to a topological phase with $\hall = 2 \times 1/2$. 
Essentially, the system is trying to approximate the rational number $c/d$ with a rational number with a smaller denominator, to find composites which are not penalized strongly by the second and the first terms in Eq.~(\ref{QJreal}) and which can condense at higher $\lambda$.
For fractions with large $c$ and $d$, a hierarchy of many topological phases will therefore be reached.
When $\eta$ is an irrational number, there will be an infinite hierarchy of topological phases, as the system tries to find a better and better rational approximation of $\eta$.
Crucially for us, the trivial insulator remains present near the $\lambda_1 = \lambda_2$ line in the window between approximately $\lambda_c/4$ and $\lambda_c$ for any $\eta$, since any topological phase requires composites containing at least two vortices.

Our $\Cnl$ property relates the phase diagram for $1/2 < \eta < 1$ to that for $0 < \eta < 1/2$.
The phase transitions are all in the same places, but topological phases with Hall conductivity $\hall$ are mapped to phases with Hall conductivity $2 - \hall$.
The trivial insulator is mapped to the BIQHE.

The exhibited phase diagrams at $\eta = 0, 1/3, 2/5$ and $1$ were obtained in Refs.~\onlinecite{FQHE,jongyeon} (panels $\eta = 0,1$ and $\eta=1/3$ are Figs.~6 and 7 in the first reference, while $\eta = 2/5$ is obtained by appropriate rescaling of Fig.~9 from the second reference
\footnote{The transition from the trivial insulator to phase where both species are superfluid at $\eta = 2/5$ was not the main focus of Ref.~\onlinecite{jongyeon} and was not determined accurately; in particular, it was determined by looking at specific heat as a function of $\lambda_1$ with $\lambda_2$ fixed.
In this paper we studied convenient ``superfluid stiffness'' (i.e., current-current correlation in convenient simulation variables) as a function of $\lambda_1 = \lambda_2$ (similar to Fig.~4 of Ref.~\onlinecite{short_range3}), obtaining a more accurate location of the transition which is plotted in Fig.~\ref{fig:fixedeta}. 
We obtain a transition at $\lambda_c \approx 0.27$, while in Ref.~\onlinecite{jongyeon} we reported a value of $\lambda_c \approx 0.23$.}
).

\subsection{$\eta = 1/2$}
As we approach $\eta = 1/2$, the width of the trivial insulator in the direction perpendicular to the $\lambda_1 = \lambda_2$ line shrinks, and exactly at $\eta = 1/2$ we do not have a trivial insulator but instead find a direct transition between the flanking superfluids where only one or the other species condenses.
The numerical phase diagram is from Ref.~\onlinecite{Loopy}, where we studied a model formulated as two species of particles with mutual $\pi$ statistics.
In the next section (following Ref.~\onlinecite{FQHE}) we will show how such a model is an exact reformulation of Eq.~(\ref{SJJ}), and also an exact reformulation of the easy-plane NCCP1 model, where a symmetry under the interchange of the two species with mutual $\pi$ statistics corresponds to the exact self-duality of the EP-NCCP1 model.\cite{Loopy, short_range3}
Here we simply state the numerical results of Ref.~\onlinecite{Loopy}, which are that there is a phase transition along the $\lambda_1 = \lambda_2$ line and that the phase transition is weakly first-order.
Note that to see the first-order nature of the transition we needed to access quite large system sizes, up to $L \geq 24$.

\subsection{Cut through the phase diagram along the self-dual line $\lambda_1 = \lambda_2$ and varying $\eta$}
\begin{figure}
\includegraphics[width=\linewidth]{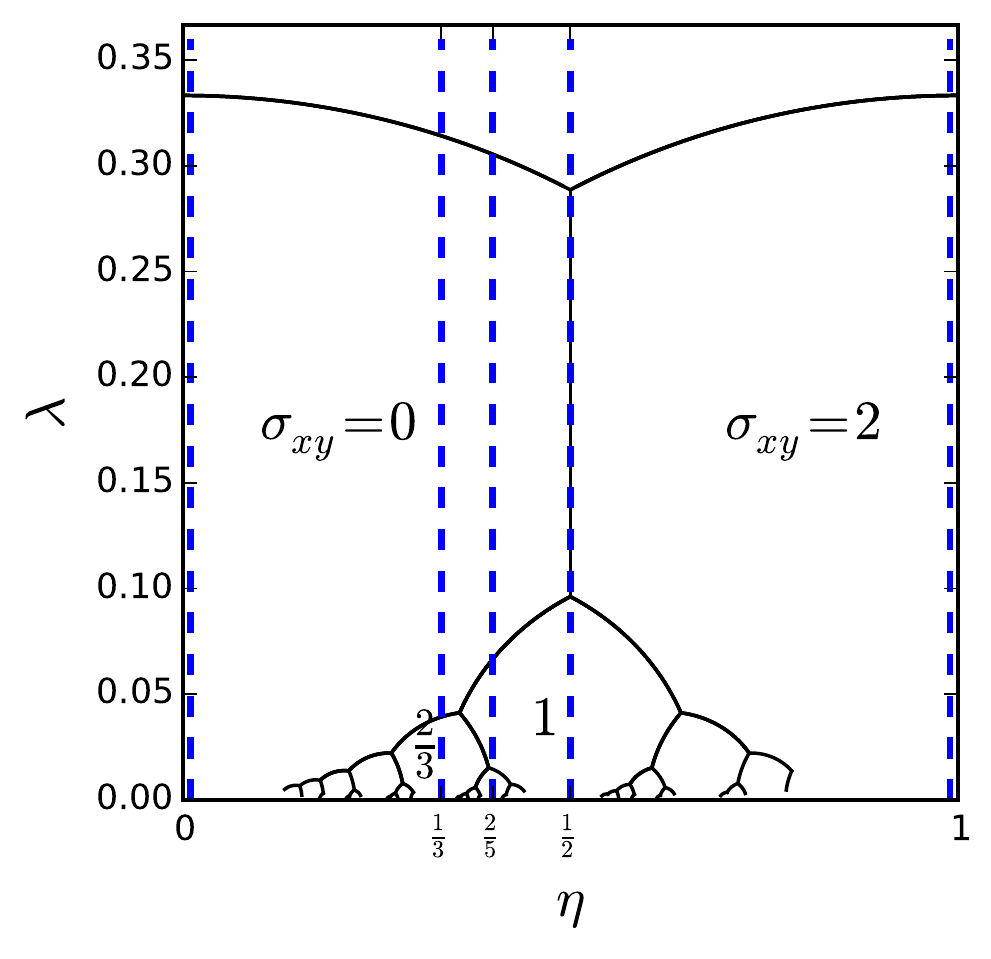}
\caption{A schematic of the phase diagram obtained by fixing $\lambda_1 = \lambda_2 = \lambda$, and varying $\eta$.
The numbers inside the phases label their Hall conductivity, with $\sigma_{xy} = 0$ the trivial phase and $\sigma_{xy} = 2$ the BIQHE.
Fractional values $\sigma_{xy} = 2c/d$ correspond to symmetry-enriched topological phases with $d$ vortices bound to $c$ bosons.}
\label{fig:self_dual}
\end{figure}

In Fig.~\ref{fig:self_dual} we present a schematic phase diagram on the ``self-dual line'' $\lambda_1 = \lambda_2 \equiv \lambda$, for all values of $\eta$ (the reason for the name ``self-dual'' will become clear in Sec.~\ref{sec:relations}).
The phase boundaries in this figure were determined analytically for a model with a somewhat different interaction than the one in this paper.
\footnote{In particular, one can take Eq.~(\ref{action}) and use interactions which are $\propto 1/k$ combined with nonlocal $w(k) = \eta'/(2\pi|f_k|^2)$.
Doing this gives the model an additional property that the interactions have the same form in terms of both $\JJ$ and $\QQ$ variables.
The combination of dualizing $\JJ_{1,2}$ to $\QQ_{1,2}$ and shifting $\eta'$ by $1$ (another property of the model derived from the fact that in the lattice loop model shifting the statistical angle by $2\pi$ does not change the contribution to the partition sum) generates the entires modular group, and this allows the phase diagram to be determined analytically.\cite{Gen2Loops}}
Obtaining a phase diagram such as Fig.~\ref{fig:self_dual} for the model in this paper cannot be done analytically, and therefore we would need to perform numerics for each value of $\eta$.
However, we expect that the topology of the phase diagrams for both models should be similar.
Indeed, if we look at the values of $\eta$ where we have performed the numerics (marked by dashed lines in Fig.~\ref{fig:fixedeta}) we see the expected structure.

An important feature of this phase diagram (and one we expect to be independent of the details of the interactions) is that when $\lambda_c/4 \lesssim \lambda \lesssim \lambda_c$ and $\eta < 1/2$, the only variables that can condense are single $\QQ_1$ and $\QQ_2$ variables, and the phase is therefore a trivial insulator.
The $\Cnl$ property tells us that in the same region of $\lambda$ but when $\eta > 1/2$, the system will be a BIQHE.
From this it follows that the existence of the $\Cnl$ symmetry implies a direct transition from the trivial insulator to the BIQHE phase when $\eta = 1/2$.

To conclude our discussion of the phase diagram, the upshot is that by requiring the species interchange symmetry (i.e., $\lambda_1 = \lambda_2$), and the nonlocal particle-hole symmetry $\Cnl$ (i.e., $\eta = 1/2$), we can place the model exactly at the transition between the trivial and BIQHE phases.

\section{Relation of the trivial-BIQHE transition to exactly self-dual EP-NCCP1 model and $\pi$-statistics model}
\label{sec:relations}

Having summarized numerical study in the specific model, we now discuss the relation of the trivial-BIQHE transition to exactly self-dual (tuned to the transition) EP-NCCP1 model and to the $\pi$-statistics model (where both species are trying to condense simultaneously).
This will allow us to see the interplay between duality and symmetry more explicitly (a topic of much recent interest), as well as come up with new models for future Monte Carlo studies.

To show that the model in Eq.~(\ref{action}) is connected to the EP-NCCP1 model we start by writing it in momentum space:
\begin{align}
S_{JJ}[\JJ_1, \JJ_2] =& \frac{1}{2} \sum_k \left[ v_1(k) |\JJ_1(k)|^2 + v_2(k) |\JJ_2(k)|^2 \right] \nonumber\\
&+ i \sum_k \theta(k) \JJ_1(-k) \cdot \vec{p}_2(k) ~, \label{SJJ}
\end{align}
where $\JJ_2 = \vec{\nabla} \times \vec{p}_2$.
Here $\theta(k)=w(k)|f_k|^2$, and the $\theta(k)$ term encodes a ``statistical interaction'' between the two loop species.
We are specifically interested in systems with short-range $v_1(r-r'), v_2(R-R'), w(R-R')$, i.e., nonsingular $v_{1, 2}(k), w(k)$ at small $k$, and hence $\theta(k) \sim k^2$ so the statistical interaction vanishes at long distances.

It is instructive to consider a version of Eq.~(\ref{QJreal}) for general potentials:\cite{FQHE,short_range3}
\begin{align}
S_{QJ}[\QQ_1, \JJ_2] = \frac{1}{2} \sum_k \Bigg[& \frac{(2\pi)^2}{v_1(k) |\vec{f}_k|^2} \left| \QQ_1(k) + \frac{\theta(k)}{2\pi} \JJ_2(k) \right|^2 \nonumber\\
& + v_2(k) |\JJ_2(k)|^2 \Bigg] ~. \label{SQJ}
\end{align}
Here the $\QQ_1$ currents have long-range interactions ($\sim 1/k^2$ in momentum space), while $\QQ_1$-$\JJ_2$ and $\JJ_2$-$\JJ_2$ interactions are short range.
[It is straightforward to check that specializing to potentials in Eqs.~(\ref{vintro}) and (\ref{tintro}) returns precisely Eq.~(\ref{QJreal})].

A simple but important reformulation of Eq.~(\ref{SQJ}) is obtained by changing variables in the partition sum as $\QQ_1 = \LL_1 + \LL_2$, $\JJ_2 = \LL_2$, which is a valid change of variables since independent summation over integer-valued $\QQ_1$ and $\JJ_2$ corresponds to independent summation over integer-valued $\LL_1$ and $\LL_2$:
\begin{align}
& S_{LL}[\LL_1, \LL_2] \equiv S_{QJ}[\LL_1 + \LL_2, \LL_2] \\
& = \frac{1}{2} \sum_k \Bigg[ \frac{(2\pi)^2}{v_1(k) |\vec{f}_k|^2} \left| \LL_1(k) + \left(1 + \frac{\theta(k)}{2\pi} \right) \LL_2(k) \right|^2 \\
& \qquad\qquad + v_2(k) |\LL_2(k)|^2 \Bigg] ~. \label{SLL}
\end{align}
Note that it is only the specific combination $\LL_1 + \LL_2$ that has long-range interactions, and we can interpret these as arising from $\LL_1$ and $\LL_2$ being coupled to the same dynamical gauge field with a generic Maxwell term.
This structure with only such a combination of loops experiencing the long-range interaction is the structure of generalized easy-plane NCCP1 model introduced in Ref.~\onlinecite{shortlight} (unlike that reference, here we are allowing the two species to have different energetics).

Applying duality transformation (e.g., as defined in Appendix A of Ref.~\onlinecite{FQHE}) to this model, denoting the variables dual to $\LL_1$ and $\LL_2$ as $\MM_1$ and $\MM_2$ respectively, we get
\begin{align}
& S_{MM}[\MM_1, \MM_2] = \frac{1}{2} \sum_k \Bigg[ v_1(k) \left| \MM_1(k) \right|^2 \label{SMM} \\
& \qquad + \frac{(2\pi)^2}{v_2(k) |\vec{f}_k|^2} \left| \MM_2(k) - \left(1 + \frac{\theta(k)}{2\pi} \right) \MM_1(k) \right|^2 \Bigg] ~. \nonumber
\end{align}
Here it is the combination $\MM_1 - \MM_2$ that has long-range interactions, and the structure of the dual theory is similar to that of the original theory up to changing sign of one of the currents.\cite{shortlight}

So far everything is completely general, except that from the outset our model $S_{JJ}$ has the symmetries Eq.~(\ref{unitaryPH}) and Eq.~(\ref{antiunitarySym}).
These basic symmetries translate to symmetries of the $S_{LL}$ model as ${\cal C}_\text{unitary}: \LL_s \to -\LL_s$ and ${\cal T}_{-+}: \LL_s \to \LL_s, i \to -i$, which are symmetries of the NCCP1 model as it was defined in Ref.~\onlinecite{shortlight}.

Let us now consider the case when the $S_{JJ}$ model has the nonlocal antiunitary particle-hole symmetry of the type discussed earlier, Eq.~(\ref{nonlocalPH}).
Since we already know that $S_{QJ}$ is real-valued, we can formulate the $\Cnl$ symmetry as invariance of $S_{QJ}$ under change of variables $\JJ_2 \to \QQ_1 - \JJ_2$ while leaving $\QQ_1$ untouched, i.e.,
\begin{align}
\Cnl ~~ &\Leftrightarrow ~~ S_{QJ}[\QQ_1, \QQ_1 - \JJ_2] = S_{QJ}[\QQ_1, \JJ_2] ~.  \\
&\Leftrightarrow ~~ S_{LL}[\LL_2, \LL_1] = S_{LL}[\LL_1, \LL_2] ~.
\label{Cnl2SLL}
\end{align}
Thus, $\Cnl$ (together with implicitly used ${\cal T}_{-+}$) is equivalent to the condition that $S_{LL}$ is invariant under interchange of $\LL_1$ and $\LL_2$.
It was such a model with the interchange symmetry that was called the easy-plane NCCP1 model in Ref.~\onlinecite{shortlight}.
It is straightforward to show that necessary and sufficient condition for this symmetry (in this class of models) is
\begin{align}
v_1(k) v_2(k) |\vec{f}_k|^2 + \theta(k)^2 + 4\pi \theta(k) = 0 ~.
\label{v1v2theta_nonlocalPH}
\end{align}
Any model of the form Eq.~(\ref{action}) whose interaction potentials satisfy the above condition is therefore equivalent to the EP-NCCP1 model.
[It is easy to check that the specific choices made in Eqs.~(\ref{vintro}) and (\ref{tintro}) at $\eta = 1/2$ satisfy this condition.]

Next, by comparing $S_{LL}$ and its dual theory $S_{MM}$, we notice that the interchange symmetry ${\cal R}$ in the boson model, Eq.~(\ref{interchange}) [i.e., $v_1(k) = v_2(k)$], implies ``exact self-duality'' of the $S_{LL}$ model in the sense that:
\begin{align}
{\cal R} ~~ \Leftrightarrow ~~ S_{MM}[\MM_1, \MM_2] = S_{LL}[\MM_2, -\MM_1] ~.
\label{R2selfduality}
\end{align}
This is why we referred to the $\lambda_1 = \lambda_2$ lines in Fig.~\ref{fig:fixedeta} (and in Refs.~\onlinecite{Loopy, short_range3}) as ``self-dual''.

Requiring both the nonlocal particle-hole symmetry and the species interchange symmetry in the original model $S_{JJ}$ thus makes the $S_{LL}$ reformulation to be the exactly-self-dual EP-NCCP1 model.
In the parameter regime where we expect a transition from $\LL_1, \LL_2$ both gapped to $\LL_1, \LL_2$ both condensed, we expect that the exactly-self-dual model lies at the transition between the two phases.

When Eq.~(\ref{v1v2theta_nonlocalPH}) is satisfied (i.e., in the presence of $\Cnl$ symmetry), we can write
\begin{align}
S_{LL}[\LL_1, \LL_2] = \frac{1}{2} \sum_k \Big[& v_+(k) |(\LL_1 + \LL_2)(k)|^2 \nonumber \\
& + v_-(k) |(\LL_1 - \LL_2)(k)|^2 \Big] ~, \label{v+v-}
\end{align}
with long-range $v_+(k) = -v_2(k) \pi/\theta(k)$ and short-range $v_-(k) = -\pi \theta(k)/[v_1(k) |\vec{f}_k|^2]$ [note that Eq.~(\ref{v1v2theta_nonlocalPH}) implies $\theta(k) < 0$ and that $v_{1,2}(k)$ and $\theta(k)$ are not independent].
If in addition we have $v_1(k) = v_2(k)$ (i.e., ${\cal R}$ symmetry), the potentials in Eq.~(\ref{v+v-}) satisfy
\begin{align}
v_+(k) v_-(k) = \frac{\pi^2}{|\vec{f}_k|^2} ~,
\label{v1v2selfdualNCCP1}
\end{align}
which is indeed a previously established condition for the exact self-duality in the EP-NCCP1 model (see, e.g., footnote~46 in Ref.~\onlinecite{short_range3}).
The model we study numerically which uses the potentials in Eqs.~(\ref{vintro}) and (\ref{tintro}) with $\eta = 1/2$ and $\lambda_1 = \lambda_2$ satisfies these conditions, but we wish to emphasize that {\it any} model satisfying Eq.~(\ref{v1v2theta_nonlocalPH}) and $v_1(k) = v_2(k)$ will be at the phase boundary between the trivial insulator and the BIQHE.

Let us consider one more reformulation of the $S_{JJ}$ model, which will allow us to make direct contact with Ref.~\onlinecite{Loopy}. 
In that paper we studied a model of the form Eq.~(\ref{SJJ}) with 
\begin{equation}
v_{G1}(k) = \frac{1}{t_1} ~, \quad v_{G2}(k) = \frac{1}{t_2} ~, \quad \theta_G(k) = \pi ~,
\label{vLoopy}
\end{equation}
i.e., model with mutual $\pi$ statistics (the subscript ``G'' will become clear below).
Using a sign-free reformulation in Ref.~\onlinecite{Loopy} which is ultimately related to the same ${\cal T}_{-+}$ symmetry of underlying short-range-interacting bosons, this model could be efficiently simulated at large sizes.
Here we use the techniques of Ref.~\onlinecite{FQHE} to show that this model is exactly related to the model described by Eqs.~(\ref{vintro}) and (\ref{tintro}), i.e., the model which is the path integral of the local Hamiltonian given in Eq.~(\ref{H}).

This reformulation uses the fact that there are two operations we can formally perform on Eq.~(\ref{SJJ}).\cite{Fradkin_SL2Z, Gen2Loops, FQHE}
One is the boson-vortex duality defined earlier, and the other is a shift of $\theta(k)$ by $2\pi$ (an operation which does not change the partition sum since $\JJ_1$, $\JJ_2$ are integer-valued).
These two operations do not commute, and in fact generate the modular group $SL(2, {\mathbb Z})$. 
Following Ref.~\onlinecite{FQHE} we can apply a modular transformation with parameters $(a, b, c, d) = (0, -1, 1, 2)$ to reformulate the model Eq.~(\ref{SJJ}) in terms of new variables $\GG_1$, $\GG_2$.
These new variables see new potentials, obtained by using Eq.~(9) from Ref.~\onlinecite{FQHE}:
\begin{align}
v_{G1/G2} = \frac{(2\pi)^2 v_{1/2}(k)}{[4\pi + \theta(k)]^2 + v_1(k) v_2(k) |\vec{f}_k|^2} ~, \label{vG}\\
\theta_G(k) = \frac{-(2\pi)^2 [4\pi + \theta(k)]}{[4\pi + \theta(k)]^2 + v_1(k) v_2(k) |\vec{f}_k|^2} ~. \label{tG}
\end{align}
Note that $\theta_G(k) \to -\pi$ for $k \to 0$ for any short-range $v_{1,2}(k)$ and $\theta(k) \sim k^2$, which is a property of the specific modular transformation used.
This formulation is particularly useful when $v_{G1/G2}$ potentials are very large and hence $\GG_1$ and $\GG_2$ particles are gapped.
[Substituting Eqs.~(\ref{vintro}) and (\ref{tintro}) with $\eta = 1/2$ into Eqs.~(\ref{vG}) and (\ref{tG}) we get precisely potentials in Eq.~(\ref{vLoopy}) with $t_1 = 4 \lambda_1$, $t_2 = 4 \lambda_2$, so this regime corresponds to small $\lambda_1$ and $\lambda_2$.
As an aside, coming from $\LL_1$ and $\LL_2$ variables with action Eq.~(\ref{v+v-}), the $\GG_1$ and $\GG_2$ variables are obtained by first changing variables to $\FF_1 \equiv \LL_2$ and $\GG_2 \equiv \LL_1 - \LL_2$ and then dualizing $\FF_1$ to $\GG_1$ without touching $\GG_2$.\cite{Loopy, short_range3}]
As shown in Ref.~\onlinecite{FQHE}, in the regime where $\GG_1$ and $\GG_2$ are gapped, the original physical bosons are in a fractionalized phase with fractional Hall conductivity $\hall = 2 \times \frac{1}{2}$, while gapped excitations $\GG_1, \GG_2$ carry fractional $1/2$ charges and have mutual $\pi$ statistics (at long distances).
We can verify that if the original boson model $S_{JJ}$ is symmetric under $\JJ_1$ and $\JJ_2$ interchange, then the $S_{GG}$ reformulation is also symmetric under $\GG_1$ and $\GG_2$ interchange: $v_1(k) = v_2(k) \implies v_{G1}(k) = v_{G2}(k)$.  

Irrespective of the interchange symmetry, if the original boson model $S_{JJ}$ has the nonlocal particle-hole symmetry $\Cnl$, then the $S_{GG}$ model has
\begin{align}
\theta_G(k) = -\pi \quad \text{for all}~k ~.
\end{align}
The statistical interaction term can then be written in real space as
$-i \pi \sum_r \GG_1(r) \cdot \vec{p}_{G2}(r)$ [where $\GG_2(R) = (\vec{\nabla} \times \vec{p}_{G2})(R)$].
On the lattice and with integer-valued $\GG_1$ and $\GG_2$ (and hence $\vec{p}_{G2}$), such $-\pi$-statistics is not distinguishable from $+\pi$ statistics.
The $S_{GG}$ model in this case has an additional structure in that
\begin{align}
\Cnl ~~ \Leftrightarrow ~~ \left( e^{-S_{GG}[\GG_1, \GG_2]} \right)^* = e^{-S_{GG}[\GG_1, \GG_2]} ~.
\end{align}
Note that this is not really a local symmetry but a statement about the ``Boltzmann weight'' calculated for the total action $S_{GG}$, and this reflects the nonlocality of the $\Cnl$ symmetry under consideration in the $S_{JJ}$ model.
Interestingly, the condition for this nonlocal symmetry has apparently a concise statement in the $\GG_1, \GG_2$ variables, unlike the original $\JJ_1, \JJ_2$ variables where we could not find a simple statement without invoking first dualizing from $\JJ_1$ to $\QQ_1$.

When we have both the species interchange and the nonlocal particle-hole symmetry in the original boson model, the $S_{GG}$ reformulation presents a model of particles with identical interactions and with mutual $\pi$ statistics.
The regime of interest here is where both particles want to condense simultaneously.
If they form a critical soup, this would correspond to a continuous transition in the EP-NCCP1 model (at the exact self-duality).
On the other hand, if they phase-separate, this corresponds to a first-order transition in the EP-NCCP1 model.
In Ref.~\onlinecite{Loopy} we studied the above question about the simultaneous condensation of both species in the specific model Eq.~(\ref{vLoopy}) and found phase-separation (i.e., first-order transition for the corresponding EP-NCCP1 model).
It would be very interesting to revisit this question in the more general family of such models that are located precisely at the potential criticality, looking for realizations where there is a criticality rather than phase separation.

We would also like to note that just having the species interchange symmetry is sufficient to produce a direct BIQHE-trivial transition.
In more general such models without explicit non-local particle-hole property, we would not know the location of the transition exactly but would need to find it by tuning some parameter.
Interestingly, in the EP-NCCP1 model, Eq.~(\ref{SLL}), this setting corresponds to requiring exact self-duality, Eq.~(\ref{R2selfduality}), without requiring interchange symmetry between its two species $\LL_1$ and $\LL_2$.
The exact self-duality guarantees that if one of the species is condensed, then the other is gapped, and vice versa.
Indeed, if, say, $\LL_1$ is condensed, then from Eq.~(\ref{R2selfduality}) we conclude that $\MM_2$ must be also condensed and hence its dual $\LL_2$ must be gapped.
Thus, if one of the species $\LL_1$ or $\LL_2$ is at a condensation transition, then so is the other.
We can then conjecture that at the transition there will be an emergent symmetry between the two species at long distances (i.e., as far as their criticality is concerned), which would then correspond to an emergent non-local particle-hole symmetry at the BIQHE-trivial transition in the original boson language.
It would be interesting to test this scenario in numerical simulations.

\section{Discussion}
\label{sec:discussion}
The nature of the transition between SPT phases and trivial insulators is an open question.
The discovery of a model with a second-order transition between these phases, especially if the model were numerically tractable such that critical exponents could be extracted, would provide data which would aid in developing a theoretical understanding of such transitions.
In this paper we have constructed an explicit model realizing direct BIQHE-trivial insulator transition and have mapped the transition to the self-dual line of the easy-plane NCCP1 model.
By connecting to earlier numerical studies of the latter,\cite{Loopy} we have shown that the transition is weakly first order.\cite{Kuklov06,Smiseth05,Kragset06,Herland2010}
Though the specific model we have studied does not have a second-order transition, we hope that our work will stimulate new searches for similar models which do.
It is encouraging that the first-order transition observed in Ref.~\onlinecite{Loopy} is quite weak, only becoming apparent for system sizes $L \geq 24$.
This leads us to hope that small modifications to our model might lead to the observation of a second-order transition.
However, the weakness of the first-order transition means that one needs to be able to access large sizes in order to accurately determine the nature of the transition.

The models described by Eq.~(\ref{H}) and studied numerically in Refs.~\onlinecite{Loopy, short_range3, FQHE, jongyeon} represent a special case of Eq.~(\ref{SJJ}) where the interactions are given by Eqs.~(\ref{vintro}) and (\ref{tintro}).
Using instead interactions of the form $v(k) \propto 1/|k|$, in Ref.~\onlinecite{Gen2Loops} we found a potential second-order such transition, though our study was limited to smaller sizes $L \leq 16$; also, these models with marginally-long-range interactions do not map to a local Hamiltonian. 
In the present paper, we have shown that any model of the form Eq.~(\ref{SJJ}) that satisfies Eq.~(\ref{v1v2theta_nonlocalPH}) and $v_1(k) = v_2(k)$ represents a transition between the BIQHE and trivial insulator.
Exploring such models with short-range interactions as well as finding new models  which can be represented by local Hamiltonians and studied numerically are interesting directions for future work.

As mentioned in the introduction, recent works established connection of the EP-NCCP1 model to QED$_3$ with two Dirac species.\cite{Senthil2006_theta, WangNahum2017, Mross2017}
The loop models in the present paper can be used to provide an exact realization of the schematic bosonized version of the $N = 2$ QED$_3$ in Ref.~\onlinecite{Mross2017}.
Specifically, we can take the $S_{JJ}$ model and formally shift $\theta(k)$ by $2\pi$; the partition sum is unchanged but the new model has bosons with mutual statistics $2\pi$ at long distances, which is the bosonized starting point in Ref.~\onlinecite{Mross2017}.
Interestingly, the nonlocal antiunitary particle-hole-like symmetry of $S_{JJ}$ corresponds to an exact self-duality condition of such bosons with $2\pi$ statistics discussed in Ref.~\onlinecite{Mross2017}, and the latter was ultimately related to the interchange symmetry of the EP-NCCP1 reformulation; this is in agreement with Eq.~(\ref{Cnl2SLL}) in the present paper directly relating $\Cnl$ and the interchange symmetry of $S_{LL}$.
We remark that the more ``mundane'' symmetries that are always present in our $S_{JJ}$ model, namely ${\cal C}_\text{local}$ and ${\cal T}_{+-}$, are both required to relate the model to the EP-NCCP1 as defined in Ref.~\onlinecite{shortlight}, and the additional ${\cal R}$ and $\Cnl$ symmetries make the EP-NCCP1 reformulation self-dual and interchange-symmetric.
All these symmetries have specific counterparts in the $N = 2$ QED$_3$, corresponding to (combinations of) fermionic time-reversal symmetry, particle-hole symmetry, fermion interchange symmetry, and exact fermionic self-duality.
Reference~\onlinecite{Mross2017} used coupled-wire approach to relate the fermionic and bosonic theories.
It would be interesting to construct such Dirac fermions directly in the present bosonic model on the lattice and pursue simulations exploring these nontrivial relations.

An interesting future direction is to study effects of disorder on the BIQHE-trivial insulator transition in this paper.
Thus, we can take the reformulation in terms of particles with mutual $\pi$ statistics, Eq.~(\ref{vLoopy}), and make $t_1$ and $t_2$ random identically-distributed variables in real space (for the study of the quantum phase transition, the couplings need to be perfectly correlated in the temporal direction).
This will maintain all the symmetries ${\cal C}_\text{unitary}$, ${\cal T}_{-+}$, $\Cnl$, and ${\cal R}$ (in probabilistic sense), and we conjecture that this would also place the model at the transition.
Another future direction is to explore bosonic models allowing noninteger particle density and nonzero external fields, which can be handled by the formalism in the present paper and which may have interesting symmetry-duality interplay as well.
For example, we can construct such models realizing BIQHE at finite boson density/field, and it would be interesting to explore transitions out of this phase.

\acknowledgements
We thank J.~Y.~Lee for previous collaboration on studies of phases and phase transitions involving bosonic SET phases.
O.I.M.~thanks D.~Mross and J.~Alicea for recent collaborations and enlightening discussions of the fermionic and bosonic dualities.
S.D.G.~is supported by Department of Energy BES Grant No.\ DE-SC0002140.
O.I.M.~is supported by National Science Foundation through Grant No.\ DMR-1619696 and also by the Institute for Quantum Information and Matter, an NSF Physics Frontiers Center, with support of the Gordon and Betty Moore Foundation.

\begin{appendix}

\section{Microscopic realization of the topologically ordered phases}
\label{app:H2topophases}

In the main text (and also in Ref.~\onlinecite{FQHE}) we argued from the path integral developed for the Hamiltonian of Eq.~(\ref{H}) that the system can realize various topologically-ordered phases.
On the other hand, the Hamiltonian language was very helpful for understanding the symmetries of the model.
In this Appendix we show how we can argue for topological phases directly from the Hamiltonian language of Eq.~(\ref{H}).
While this material is tangential to the subject of the main text, we hope to provide more examples of Hamiltonian thinking while at the same time make our original model realization of the BIQHE and the fractionalized SET cousins more useful to a much broader readership familiar with exactly solvable models for topological phases.\cite{Kitaev2003}

We will only consider the simplest case $\eta = 1/d$, which can represent the BIQHE phase at $d = 1$ or fractionalized phases for $d > 1$.
Let us also consider the limit $\kappa_{1, 2} = 0$, which in the path integral language corresponds to $\lambda_{1, 2} = 0$ and therefore puts the model deep in the topological phase with $\hall = 2 \times 1/d$ discussed in Sec.~\ref{subsec:fract_eta}.
[Strictly speaking, the loop model in Eq.~(\ref{action}) corresponds to the $h_{1, 2} \to \infty$ limit, but we will not require this in the direct analysis below.]
In this case, we observe that the remaining terms in the Hamiltonian $H_{h1}$, $H_{h2}$, $H_{u1}$, and $H_{u2}$ all commute.
We can then define ground state manifold by requirement of simultaneously minimizing all these terms:
\begin{align}
& e^{i \left[\nabla_j \hat{\phi}_1(\br) - \sqrt{2\pi d}\, \hat{\alpha}_{1j}(\br) \right]} =
  e^{i \left[\nabla_j \hat{\phi}_2(\bR) - \sqrt{2\pi d}\, \hat{\alpha}_{2j}(\bR) \right]} = 1 ~, \label{minHh} \\
& \hat{n}_1(\br) + \frac{\hat{B}_2(\br)}{\sqrt{2\pi d}} \, = \,
  \hat{n}_2(\bR) + \frac{\hat{B}_1(\bR)}{\sqrt{2\pi d}} \, = \, 0 ~. \label{minHu} 
\end{align}
These equalities are understood as specifying operator eigenvalues in the ground state manifold.
The first line is for each direct lattice link $\br, j$ and each dual lattice link $\bR, j$, while the second line is for each direct lattice site $\br$ and each dual lattice site $\bR$.
In the second line, we defined ``flux'' variables
\begin{align}
\hat{B}_1(\bR) \equiv ({\bm \nabla} \wedge \hat{\bm \alpha}_1)(\bR) ~, \quad
\hat{B}_2(\br) \equiv ({\bm \nabla} \wedge \hat{\bm \alpha}_2)(\br) ~.
\end{align}
It is easy to see that these requirements on the ground state manifold are consistent.
Indeed, Eq.~(\ref{minHh}) implies $\sqrt{2\pi d}\, B_{1,2} = -2\pi q_{1,2}$ with integer-valued $q_{1,2}$ (the sign is chosen just for convenience).
The combinations in Eq.~(\ref{minHu}) become $n_{1,2} - q_{2,1} / d$ and can be made zero by requiring $q_{2,1} = d n_{1,2}$.
The physics is that $q_{1,2}$ are vorticities of the boson fields, and the Hamiltonian binds $d$ vortices of one species to a single charge of the other species.

A more detailed construction of ground states that indeed satisfy Eqs.~(\ref{minHh}) and (\ref{minHu}) will be given below.
However, we can already demonstrate ground state degeneracy (for $d > 1$) when the system is placed on a two-dimensional torus.
Assuming $L_x \times L_y$ torus with direct lattice sites labeled by integer coordinates, $0 \leq x \leq L_x - 1$, $0 \leq y \leq L_y - 1$, and dual lattice sites labeled by half-integer coordinates, we define operators
\begin{align*}
& \hat{C}_{1x} \equiv \sum_{x=0}^{L_x - 1} \hat{\alpha}_{1x}(x, 0) ~, \quad \hat{C}_{2x} \equiv \sum_{x=0}^{L_x - 1} \hat{\alpha}_{2x} \left(x + \frac{1}{2}, \frac{1}{2} \right) ~, \\
& \hat{C}_{1y} \equiv \sum_{y=0}^{L_y - 1} \hat{\alpha}_{1y}(0, y) ~, \quad \hat{C}_{2y} \equiv \sum_{y=0}^{L_y - 1} \hat{\alpha}_{2y} \left(\frac{1}{2}, y + \frac{1}{2} \right) ~.
\end{align*}
$\hat{C}_{sj}$ is a ``circulation'' of the field $\hat{\alpha}_{sj}$ along a fixed line around the torus in the $\hat{j}$ direction.
It is easy to check that operators
\begin{align}
\hat{W}_{sj} \equiv e^{i \sqrt{\frac{2\pi}{d}} \hat{C}_{sj}}
\end{align}
commute with the Hamiltonian Eq.~(\ref{H}) when $\kappa_{1, 2} = 0$ and hence preserve the ground state manifold defined by Eqs.~(\ref{minHh}) and (\ref{minHu}).
However, these operators do not commute with each other; specifically,
\begin{align}
\hat{W}_{1x} \hat{W}_{2y} = \hat{W}_{2y} \hat{W}_{1x} e^{i \frac{2\pi}{d}} ~, \quad 
\hat{W}_{2x} \hat{W}_{1y} = \hat{W}_{1y} \hat{W}_{2x} e^{i \frac{2\pi}{d}} ~.
\end{align}
Such commutators immediately imply $d^2$-fold degeneracy and therefore topological order, similar to ${\mathbb Z}_d$ toric code.\cite{Kitaev2003}

To produce explicit ground state wavefunctions, we first note that $L_x L_y - 1$ operators $\hat{B}_1(\bR)$ [e.g., requiring $\bR \neq (1/2,1/2)$] are linearly independent combinations of $\hat{\alpha}_{1j}(\br)$ and are linearly independent with $\hat{C}_{1x}$.
Similarly, $L_x L_y - 1$ operators $\hat{B}_2(\br)$ [e.g, requiring $\br \neq (0, 0)$] and operator $\hat{C}_{2x}$ are linearly independent combinations of $\hat{\alpha}_{2j}(\bR)$.
Furthermore, all these operators commute with each other.
This allows us to write a complete basis of the Hilbert space of the $2 L_x L_y$ oscillators as eigenbasis of the above operators with independent eigenvalues.
Using this basis for the oscillators and the number basis for the bosons, we can construct ground state wavefunctions as superpositions of states specified by $n_1(\br), n_2(\bR), B_1[\bR \neq (1/2,1/2)] = -\sqrt{2\pi d} \, n_2(\bR), B_2[\br \neq (0,0)] = -\sqrt{2\pi d} \, n_1(\br), C_{1x} = \sqrt{2\pi/d} \, m_{1x}, C_{2x} = \sqrt{2\pi/d} \, m_{2x}$.
Here $m_{1x}$ and $m_{2x}$ are integers independent of the other variables, and the conditions on $C_{1x}$ and $C_{2x}$ follow by multiplying constraints Eq.~(\ref{minHh}) along the corresponding loops around the torus.
Now we can show that the boson hopping operators in $H_{h1, h2}$ [i.e., operators in Eq.~(\ref{minHh})] can connect all of the above states specified by independent $n_1(\br), n_2(\bR), m_{1x}, m_{2x}$ with each other, except that only $m_{1x}$ differing by a multiple of $d$ are connected, and similarly for $m_{2x}$.
[Here we also assumed fixed total rotor numbers $\sum_\br n_1(\br) = \sum_\bR n_2(\bR) = 0$ appropriate for bosons at integer density.]
We can hence construct $d^2$ independent ground state wavefunctions which are superpositions (with in general complex but uniquely fixed amplitudes with equal absolute values) of the connected states, in agreement with the operator argument for the topological degeneracy given earlier.

Just like in the ${\mathbb Z}_d$ toric code, we can consider open-string counterparts of $\hat{W}_1$ and $\hat{W}_2$ operators; schematically,
\begin{align}
e^{i \sqrt{\frac{2\pi}{d}} \int^{\br} \hat{\bm \alpha}_1 \cdot d{\bm l}} ~~\text{and}~~
e^{i \sqrt{\frac{2\pi}{d}} \int^{\bR} \hat{\bm \alpha}_2 \cdot d{\bm l}} ~,
\end{align}
where we focus on just one end of the string.
It is easy to see that these operators create excitations residing on the direct and dual lattices respectively, with energies $\frac{u_1}{2} \frac{1}{d^2}$ and $\frac{u_2}{2} \frac{1}{d^2}$, and that these excitations have mutual statistics of $2\pi/d$.
Furthermore, these excitations carry fractional charges $1/d$ of the first and second $U(1)$ symmetries respectively.
Indeed, the first operator raised to power $d$ when applied to a ground state is, by Eq.~(\ref{minHh}), identical to acting with $e^{i \phi_1({\br})}$ on the same state, which adds charge 1 of the first boson species; hence the first operator acting once must add fractional charge $1/d$.
The obtained description of the excitations of the topological phase is in agreement with our conclusions in Ref.~\onlinecite{FQHE} using path integral approach.
\end{appendix}

\bibliography{thesis.bib}

\begin{thebibliography}{53}%
\makeatletter
\providecommand \@ifxundefined [1]{%
 \@ifx{#1\undefined}
}%
\providecommand \@ifnum [1]{%
 \ifnum #1\expandafter \@firstoftwo
 \else \expandafter \@secondoftwo
 \fi
}%
\providecommand \@ifx [1]{%
 \ifx #1\expandafter \@firstoftwo
 \else \expandafter \@secondoftwo
 \fi
}%
\providecommand \natexlab [1]{#1}%
\providecommand \enquote  [1]{``#1''}%
\providecommand \bibnamefont  [1]{#1}%
\providecommand \bibfnamefont [1]{#1}%
\providecommand \citenamefont [1]{#1}%
\providecommand \href@noop [0]{\@secondoftwo}%
\providecommand \href [0]{\begingroup \@sanitize@url \@href}%
\providecommand \@href[1]{\@@startlink{#1}\@@href}%
\providecommand \@@href[1]{\endgroup#1\@@endlink}%
\providecommand \@sanitize@url [0]{\catcode `\\12\catcode `\$12\catcode
  `\&12\catcode `\#12\catcode `\^12\catcode `\_12\catcode `\%12\relax}%
\providecommand \@@startlink[1]{}%
\providecommand \@@endlink[0]{}%
\providecommand \url  [0]{\begingroup\@sanitize@url \@url }%
\providecommand \@url [1]{\endgroup\@href {#1}{\urlprefix }}%
\providecommand \urlprefix  [0]{URL }%
\providecommand \Eprint [0]{\href }%
\providecommand \doibase [0]{http://dx.doi.org/}%
\providecommand \selectlanguage [0]{\@gobble}%
\providecommand \bibinfo  [0]{\@secondoftwo}%
\providecommand \bibfield  [0]{\@secondoftwo}%
\providecommand \translation [1]{[#1]}%
\providecommand \BibitemOpen [0]{}%
\providecommand \bibitemStop [0]{}%
\providecommand \bibitemNoStop [0]{.\EOS\space}%
\providecommand \EOS [0]{\spacefactor3000\relax}%
\providecommand \BibitemShut  [1]{\csname bibitem#1\endcsname}%
\let\auto@bib@innerbib\@empty
\bibitem [{\citenamefont {Prange}\ and\ \citenamefont
  {Girvin}(1990)}]{PrangeGirvin}%
  \BibitemOpen
  \bibfield  {author} {\bibinfo {author} {\bibfnamefont {R.~E.}\ \bibnamefont
  {Prange}}\ and\ \bibinfo {author} {\bibfnamefont {S.~M.}\ \bibnamefont
  {Girvin}},\ }\href@noop {} {\emph {\bibinfo {title} {The Quantum Hall
  Effect}}}\ (\bibinfo  {publisher} {Springer-Verlag, New York},\ \bibinfo
  {year} {1990})\BibitemShut {NoStop}%
\bibitem [{\citenamefont {Sarma}\ and\ \citenamefont
  {Pinczuk}(1996)}]{DasSarmaBook}%
  \BibitemOpen
  \bibfield  {author} {\bibinfo {author} {\bibfnamefont {S.~D.}\ \bibnamefont
  {Sarma}}\ and\ \bibinfo {author} {\bibfnamefont {A.}~\bibnamefont
  {Pinczuk}},\ }\href@noop {} {\emph {\bibinfo {title} {{Perspectives in
  quantum Hall effects : novel quantum liquids in low-dimensional semiconductor
  structures}}}}\ (\bibinfo  {publisher} {Wiley},\ \bibinfo {address}
  {Chichester},\ \bibinfo {year} {1996})\BibitemShut {NoStop}%
\bibitem [{\citenamefont {Wen}(2007)}]{Wen_book}%
  \BibitemOpen
  \bibfield  {author} {\bibinfo {author} {\bibfnamefont {X.-G.}\ \bibnamefont
  {Wen}},\ }\href@noop {} {\emph {\bibinfo {title} {Quantum field theory of
  many-body systems: from the origin of sound to an origin of light and
  electrons}}}\ (\bibinfo  {publisher} {Oxford University Press},\ \bibinfo
  {address} {Oxford},\ \bibinfo {year} {2007})\BibitemShut {NoStop}%
\bibitem [{\citenamefont {Chen}\ \emph {et~al.}(2012)\citenamefont {Chen},
  \citenamefont {Gu}, \citenamefont {Liu},\ and\ \citenamefont
  {Wen}}]{Chen2012_Science}%
  \BibitemOpen
  \bibfield  {author} {\bibinfo {author} {\bibfnamefont {X.}~\bibnamefont
  {Chen}}, \bibinfo {author} {\bibfnamefont {Z.-C.}\ \bibnamefont {Gu}},
  \bibinfo {author} {\bibfnamefont {Z.-X.}\ \bibnamefont {Liu}}, \ and\
  \bibinfo {author} {\bibfnamefont {X.-G.}\ \bibnamefont {Wen}},\ }\href
  {\doibase 10.1126/science.1227224} {\bibfield  {journal} {\bibinfo  {journal}
  {Science}\ }\textbf {\bibinfo {volume} {338}},\ \bibinfo {pages} {1604}
  (\bibinfo {year} {2012})}\BibitemShut {NoStop}%
\bibitem [{\citenamefont {Chen}\ \emph {et~al.}(2013)\citenamefont {Chen},
  \citenamefont {Gu}, \citenamefont {Liu},\ and\ \citenamefont
  {Wen}}]{Chen2013_PRB}%
  \BibitemOpen
  \bibfield  {author} {\bibinfo {author} {\bibfnamefont {X.}~\bibnamefont
  {Chen}}, \bibinfo {author} {\bibfnamefont {Z.-C.}\ \bibnamefont {Gu}},
  \bibinfo {author} {\bibfnamefont {Z.-X.}\ \bibnamefont {Liu}}, \ and\
  \bibinfo {author} {\bibfnamefont {X.-G.}\ \bibnamefont {Wen}},\ }\href
  {\doibase 10.1103/PhysRevB.87.155114} {\bibfield  {journal} {\bibinfo
  {journal} {Phys. Rev. B}\ }\textbf {\bibinfo {volume} {87}},\ \bibinfo
  {pages} {155114} (\bibinfo {year} {2013})}\BibitemShut {NoStop}%
\bibitem [{\citenamefont {Senthil}(2015)}]{Senthil_SPTreview2015}%
  \BibitemOpen
  \bibfield  {author} {\bibinfo {author} {\bibfnamefont {T.}~\bibnamefont
  {Senthil}},\ }\href {\doibase 10.1146/annurev-conmatphys-031214-014740}
  {\bibfield  {journal} {\bibinfo  {journal} {Annual Review of Condensed Matter
  Physics}\ }\textbf {\bibinfo {volume} {6}},\ \bibinfo {pages} {299} (\bibinfo
  {year} {2015})}\BibitemShut {NoStop}%
\bibitem [{\citenamefont {Gu}\ and\ \citenamefont {Wen}(2009)}]{Gu2009}%
  \BibitemOpen
  \bibfield  {author} {\bibinfo {author} {\bibfnamefont {Z.-C.}\ \bibnamefont
  {Gu}}\ and\ \bibinfo {author} {\bibfnamefont {X.-G.}\ \bibnamefont {Wen}},\
  }\href {\doibase 10.1103/PhysRevB.80.155131} {\bibfield  {journal} {\bibinfo
  {journal} {Phys. Rev. B}\ }\textbf {\bibinfo {volume} {80}},\ \bibinfo
  {pages} {155131} (\bibinfo {year} {2009})}\BibitemShut {NoStop}%
\bibitem [{\citenamefont {Jiang}\ \emph {et~al.}(2010)\citenamefont {Jiang},
  \citenamefont {Rachel}, \citenamefont {Weng}, \citenamefont {Zhang},\ and\
  \citenamefont {Wang}}]{Jiang2010}%
  \BibitemOpen
  \bibfield  {author} {\bibinfo {author} {\bibfnamefont {H.-C.}\ \bibnamefont
  {Jiang}}, \bibinfo {author} {\bibfnamefont {S.}~\bibnamefont {Rachel}},
  \bibinfo {author} {\bibfnamefont {Z.-Y.}\ \bibnamefont {Weng}}, \bibinfo
  {author} {\bibfnamefont {S.-C.}\ \bibnamefont {Zhang}}, \ and\ \bibinfo
  {author} {\bibfnamefont {Z.}~\bibnamefont {Wang}},\ }\href {\doibase
  10.1103/PhysRevB.82.220403} {\bibfield  {journal} {\bibinfo  {journal} {Phys.
  Rev. B}\ }\textbf {\bibinfo {volume} {82}},\ \bibinfo {pages} {220403}
  (\bibinfo {year} {2010})}\BibitemShut {NoStop}%
\bibitem [{\citenamefont {Lange}\ \emph {et~al.}(2015)\citenamefont {Lange},
  \citenamefont {Ejima},\ and\ \citenamefont {Fehske}}]{Lange2015}%
  \BibitemOpen
  \bibfield  {author} {\bibinfo {author} {\bibfnamefont {F.}~\bibnamefont
  {Lange}}, \bibinfo {author} {\bibfnamefont {S.}~\bibnamefont {Ejima}}, \ and\
  \bibinfo {author} {\bibfnamefont {H.}~\bibnamefont {Fehske}},\ }\href
  {\doibase 10.1103/PhysRevB.92.041120} {\bibfield  {journal} {\bibinfo
  {journal} {Phys. Rev. B}\ }\textbf {\bibinfo {volume} {92}},\ \bibinfo
  {pages} {041120} (\bibinfo {year} {2015})}\BibitemShut {NoStop}%
\bibitem [{\citenamefont {Lu}\ and\ \citenamefont
  {Vishwanath}(2012)}]{LuVishwanath2012}%
  \BibitemOpen
  \bibfield  {author} {\bibinfo {author} {\bibfnamefont {Y.-M.}\ \bibnamefont
  {Lu}}\ and\ \bibinfo {author} {\bibfnamefont {A.}~\bibnamefont
  {Vishwanath}},\ }\href {\doibase 10.1103/PhysRevB.86.125119} {\bibfield
  {journal} {\bibinfo  {journal} {Phys. Rev. B}\ }\textbf {\bibinfo {volume}
  {86}},\ \bibinfo {pages} {125119} (\bibinfo {year} {2012})}\BibitemShut
  {NoStop}%
\bibitem [{\citenamefont {Senthil}\ and\ \citenamefont
  {Levin}(2013)}]{SenthilLevin2012}%
  \BibitemOpen
  \bibfield  {author} {\bibinfo {author} {\bibfnamefont {T.}~\bibnamefont
  {Senthil}}\ and\ \bibinfo {author} {\bibfnamefont {M.}~\bibnamefont
  {Levin}},\ }\href {\doibase 10.1103/PhysRevLett.110.046801} {\bibfield
  {journal} {\bibinfo  {journal} {Phys. Rev. Lett.}\ }\textbf {\bibinfo
  {volume} {110}},\ \bibinfo {pages} {046801} (\bibinfo {year}
  {2013})}\BibitemShut {NoStop}%
\bibitem [{\citenamefont {Geraedts}\ and\ \citenamefont
  {Motrunich}(2013)}]{FQHE}%
  \BibitemOpen
  \bibfield  {author} {\bibinfo {author} {\bibfnamefont {S.~D.}\ \bibnamefont
  {Geraedts}}\ and\ \bibinfo {author} {\bibfnamefont {O.~I.}\ \bibnamefont
  {Motrunich}},\ }\href {\doibase http://dx.doi.org/10.1016/j.aop.2013.03.017}
  {\bibfield  {journal} {\bibinfo  {journal} {Annals of Physics}\ }\textbf
  {\bibinfo {volume} {334}},\ \bibinfo {pages} {288 } (\bibinfo {year}
  {2013})}\BibitemShut {NoStop}%
\bibitem [{\citenamefont {Furukawa}\ and\ \citenamefont
  {Ueda}(2013)}]{Furukawa-PhysRevLett.111.090401}%
  \BibitemOpen
  \bibfield  {author} {\bibinfo {author} {\bibfnamefont {S.}~\bibnamefont
  {Furukawa}}\ and\ \bibinfo {author} {\bibfnamefont {M.}~\bibnamefont
  {Ueda}},\ }\href {\doibase 10.1103/PhysRevLett.111.090401} {\bibfield
  {journal} {\bibinfo  {journal} {Phys. Rev. Lett.}\ }\textbf {\bibinfo
  {volume} {111}},\ \bibinfo {pages} {090401} (\bibinfo {year}
  {2013})}\BibitemShut {NoStop}%
\bibitem [{\citenamefont {Wu}\ and\ \citenamefont
  {Jain}(2013)}]{Wu-PhysRevB.87.245123}%
  \BibitemOpen
  \bibfield  {author} {\bibinfo {author} {\bibfnamefont {Y.-H.}\ \bibnamefont
  {Wu}}\ and\ \bibinfo {author} {\bibfnamefont {J.~K.}\ \bibnamefont {Jain}},\
  }\href {\doibase 10.1103/PhysRevB.87.245123} {\bibfield  {journal} {\bibinfo
  {journal} {Phys. Rev. B}\ }\textbf {\bibinfo {volume} {87}},\ \bibinfo
  {pages} {245123} (\bibinfo {year} {2013})}\BibitemShut {NoStop}%
\bibitem [{\citenamefont {Regnault}\ and\ \citenamefont
  {Senthil}(2013)}]{Regnault-PhysRevB.88.161106}%
  \BibitemOpen
  \bibfield  {author} {\bibinfo {author} {\bibfnamefont {N.}~\bibnamefont
  {Regnault}}\ and\ \bibinfo {author} {\bibfnamefont {T.}~\bibnamefont
  {Senthil}},\ }\href {\doibase 10.1103/PhysRevB.88.161106} {\bibfield
  {journal} {\bibinfo  {journal} {Phys. Rev. B}\ }\textbf {\bibinfo {volume}
  {88}},\ \bibinfo {pages} {161106} (\bibinfo {year} {2013})}\BibitemShut
  {NoStop}%
\bibitem [{\citenamefont {Geraedts}\ \emph {et~al.}(2017)\citenamefont
  {Geraedts}, \citenamefont {Repellin}, \citenamefont {Wang}, \citenamefont
  {Mong}, \citenamefont {Senthil},\ and\ \citenamefont
  {Regnault}}]{Geraedts2017}%
  \BibitemOpen
  \bibfield  {author} {\bibinfo {author} {\bibfnamefont {S.~D.}\ \bibnamefont
  {Geraedts}}, \bibinfo {author} {\bibfnamefont {C.}~\bibnamefont {Repellin}},
  \bibinfo {author} {\bibfnamefont {C.}~\bibnamefont {Wang}}, \bibinfo {author}
  {\bibfnamefont {R.~S.~K.}\ \bibnamefont {Mong}}, \bibinfo {author}
  {\bibfnamefont {T.}~\bibnamefont {Senthil}}, \ and\ \bibinfo {author}
  {\bibfnamefont {N.}~\bibnamefont {Regnault}},\ }\href@noop {} {\bibfield
  {journal} {\bibinfo  {journal} {arXiv:cond-mat}\ ,\ \bibinfo {pages}
  {1704.01594}} (\bibinfo {year} {2017})}\BibitemShut {NoStop}%
\bibitem [{\citenamefont {M{\"o}ller}\ and\ \citenamefont
  {Cooper}(2009)}]{Moller:2009p184}%
  \BibitemOpen
  \bibfield  {author} {\bibinfo {author} {\bibfnamefont {G.}~\bibnamefont
  {M{\"o}ller}}\ and\ \bibinfo {author} {\bibfnamefont {N.~R.}\ \bibnamefont
  {Cooper}},\ }\href {\doibase 10.1103/PhysRevLett.103.105303} {\bibfield
  {journal} {\bibinfo  {journal} {Phys. Rev. Lett.}\ }\textbf {\bibinfo
  {volume} {103}},\ \bibinfo {pages} {105303} (\bibinfo {year}
  {2009})}\BibitemShut {NoStop}%
\bibitem [{\citenamefont {Sterdyniak}\ \emph {et~al.}(2015)\citenamefont
  {Sterdyniak}, \citenamefont {Cooper},\ and\ \citenamefont
  {Regnault}}]{Sterdyniak-PhysRevLett.115.116802}%
  \BibitemOpen
  \bibfield  {author} {\bibinfo {author} {\bibfnamefont {A.}~\bibnamefont
  {Sterdyniak}}, \bibinfo {author} {\bibfnamefont {N.~R.}\ \bibnamefont
  {Cooper}}, \ and\ \bibinfo {author} {\bibfnamefont {N.}~\bibnamefont
  {Regnault}},\ }\href {\doibase 10.1103/PhysRevLett.115.116802} {\bibfield
  {journal} {\bibinfo  {journal} {Phys. Rev. Lett.}\ }\textbf {\bibinfo
  {volume} {115}},\ \bibinfo {pages} {116802} (\bibinfo {year}
  {2015})}\BibitemShut {NoStop}%
\bibitem [{\citenamefont {He}\ \emph {et~al.}(2015)\citenamefont {He},
  \citenamefont {Bhattacharjee}, \citenamefont {Moessner},\ and\ \citenamefont
  {Pollmann}}]{He2015}%
  \BibitemOpen
  \bibfield  {author} {\bibinfo {author} {\bibfnamefont {Y.-C.}\ \bibnamefont
  {He}}, \bibinfo {author} {\bibfnamefont {S.}~\bibnamefont {Bhattacharjee}},
  \bibinfo {author} {\bibfnamefont {R.}~\bibnamefont {Moessner}}, \ and\
  \bibinfo {author} {\bibfnamefont {F.}~\bibnamefont {Pollmann}},\ }\href
  {\doibase 10.1103/PhysRevLett.115.116803} {\bibfield  {journal} {\bibinfo
  {journal} {Phys. Rev. Lett.}\ }\textbf {\bibinfo {volume} {115}},\ \bibinfo
  {pages} {116803} (\bibinfo {year} {2015})}\BibitemShut {NoStop}%
\bibitem [{\citenamefont {Zeng}\ \emph {et~al.}(2016)\citenamefont {Zeng},
  \citenamefont {Zhu},\ and\ \citenamefont {Sheng}}]{Zeng-PhysRevB.93.195121}%
  \BibitemOpen
  \bibfield  {author} {\bibinfo {author} {\bibfnamefont {T.-S.}\ \bibnamefont
  {Zeng}}, \bibinfo {author} {\bibfnamefont {W.}~\bibnamefont {Zhu}}, \ and\
  \bibinfo {author} {\bibfnamefont {D.~N.}\ \bibnamefont {Sheng}},\ }\href
  {\doibase 10.1103/PhysRevB.93.195121} {\bibfield  {journal} {\bibinfo
  {journal} {Phys. Rev. B}\ }\textbf {\bibinfo {volume} {93}},\ \bibinfo
  {pages} {195121} (\bibinfo {year} {2016})}\BibitemShut {NoStop}%
\bibitem [{\citenamefont {Grover}\ and\ \citenamefont
  {Vishwanath}(2013)}]{GroverVishwanath2013}%
  \BibitemOpen
  \bibfield  {author} {\bibinfo {author} {\bibfnamefont {T.}~\bibnamefont
  {Grover}}\ and\ \bibinfo {author} {\bibfnamefont {A.}~\bibnamefont
  {Vishwanath}},\ }\href {\doibase 10.1103/PhysRevB.87.045129} {\bibfield
  {journal} {\bibinfo  {journal} {Phys. Rev. B}\ }\textbf {\bibinfo {volume}
  {87}},\ \bibinfo {pages} {045129} (\bibinfo {year} {2013})}\BibitemShut
  {NoStop}%
\bibitem [{\citenamefont {Lu}\ and\ \citenamefont {Lee}(2014)}]{LuLee2014_QPT}%
  \BibitemOpen
  \bibfield  {author} {\bibinfo {author} {\bibfnamefont {Y.-M.}\ \bibnamefont
  {Lu}}\ and\ \bibinfo {author} {\bibfnamefont {D.-H.}\ \bibnamefont {Lee}},\
  }\href {\doibase 10.1103/PhysRevB.89.195143} {\bibfield  {journal} {\bibinfo
  {journal} {Phys. Rev. B}\ }\textbf {\bibinfo {volume} {89}},\ \bibinfo
  {pages} {195143} (\bibinfo {year} {2014})}\BibitemShut {NoStop}%
\bibitem [{\citenamefont {Slagle}\ \emph {et~al.}(2015)\citenamefont {Slagle},
  \citenamefont {You},\ and\ \citenamefont {Xu}}]{SlagleYouXu2015}%
  \BibitemOpen
  \bibfield  {author} {\bibinfo {author} {\bibfnamefont {K.}~\bibnamefont
  {Slagle}}, \bibinfo {author} {\bibfnamefont {Y.-Z.}\ \bibnamefont {You}}, \
  and\ \bibinfo {author} {\bibfnamefont {C.}~\bibnamefont {Xu}},\ }\href
  {\doibase 10.1103/PhysRevB.91.115121} {\bibfield  {journal} {\bibinfo
  {journal} {Phys. Rev. B}\ }\textbf {\bibinfo {volume} {91}},\ \bibinfo
  {pages} {115121} (\bibinfo {year} {2015})}\BibitemShut {NoStop}%
\bibitem [{\citenamefont {He}\ \emph {et~al.}(2016)\citenamefont {He},
  \citenamefont {Wu}, \citenamefont {You}, \citenamefont {Xu}, \citenamefont
  {Meng},\ and\ \citenamefont {Lu}}]{HeWuYouXuMengLu2016}%
  \BibitemOpen
  \bibfield  {author} {\bibinfo {author} {\bibfnamefont {Y.-Y.}\ \bibnamefont
  {He}}, \bibinfo {author} {\bibfnamefont {H.-Q.}\ \bibnamefont {Wu}}, \bibinfo
  {author} {\bibfnamefont {Y.-Z.}\ \bibnamefont {You}}, \bibinfo {author}
  {\bibfnamefont {C.}~\bibnamefont {Xu}}, \bibinfo {author} {\bibfnamefont
  {Z.~Y.}\ \bibnamefont {Meng}}, \ and\ \bibinfo {author} {\bibfnamefont
  {Z.-Y.}\ \bibnamefont {Lu}},\ }\href {\doibase 10.1103/PhysRevB.93.115150}
  {\bibfield  {journal} {\bibinfo  {journal} {Phys. Rev. B}\ }\textbf {\bibinfo
  {volume} {93}},\ \bibinfo {pages} {115150} (\bibinfo {year}
  {2016})}\BibitemShut {NoStop}%
\bibitem [{\citenamefont {Bi}\ \emph {et~al.}(2017)\citenamefont {Bi},
  \citenamefont {Zhang}, \citenamefont {You}, \citenamefont {Young},
  \citenamefont {Balents}, \citenamefont {Liu},\ and\ \citenamefont
  {Xu}}]{BiZhangYouYoungBalentsLiuXu2017}%
  \BibitemOpen
  \bibfield  {author} {\bibinfo {author} {\bibfnamefont {Z.}~\bibnamefont
  {Bi}}, \bibinfo {author} {\bibfnamefont {R.}~\bibnamefont {Zhang}}, \bibinfo
  {author} {\bibfnamefont {Y.-Z.}\ \bibnamefont {You}}, \bibinfo {author}
  {\bibfnamefont {A.}~\bibnamefont {Young}}, \bibinfo {author} {\bibfnamefont
  {L.}~\bibnamefont {Balents}}, \bibinfo {author} {\bibfnamefont {C.-X.}\
  \bibnamefont {Liu}}, \ and\ \bibinfo {author} {\bibfnamefont
  {C.}~\bibnamefont {Xu}},\ }\href {\doibase 10.1103/PhysRevLett.118.126801}
  {\bibfield  {journal} {\bibinfo  {journal} {Phys. Rev. Lett.}\ }\textbf
  {\bibinfo {volume} {118}},\ \bibinfo {pages} {126801} (\bibinfo {year}
  {2017})}\BibitemShut {NoStop}%
\bibitem [{\citenamefont {Fuji}\ \emph {et~al.}(2016)\citenamefont {Fuji},
  \citenamefont {He}, \citenamefont {Bhattacharjee},\ and\ \citenamefont
  {Pollmann}}]{FujiHe2016}%
  \BibitemOpen
  \bibfield  {author} {\bibinfo {author} {\bibfnamefont {Y.}~\bibnamefont
  {Fuji}}, \bibinfo {author} {\bibfnamefont {Y.-C.}\ \bibnamefont {He}},
  \bibinfo {author} {\bibfnamefont {S.}~\bibnamefont {Bhattacharjee}}, \ and\
  \bibinfo {author} {\bibfnamefont {F.}~\bibnamefont {Pollmann}},\ }\href
  {\doibase 10.1103/PhysRevB.93.195143} {\bibfield  {journal} {\bibinfo
  {journal} {Phys. Rev. B}\ }\textbf {\bibinfo {volume} {93}},\ \bibinfo
  {pages} {195143} (\bibinfo {year} {2016})}\BibitemShut {NoStop}%
\bibitem [{\citenamefont {Geraedts}\ and\ \citenamefont
  {Motrunich}(2012{\natexlab{a}})}]{Loopy}%
  \BibitemOpen
  \bibfield  {author} {\bibinfo {author} {\bibfnamefont {S.~D.}\ \bibnamefont
  {Geraedts}}\ and\ \bibinfo {author} {\bibfnamefont {O.~I.}\ \bibnamefont
  {Motrunich}},\ }\href {\doibase 10.1103/PhysRevB.85.045114} {\bibfield
  {journal} {\bibinfo  {journal} {Phys. Rev. B}\ }\textbf {\bibinfo {volume}
  {85}},\ \bibinfo {pages} {045114} (\bibinfo {year}
  {2012}{\natexlab{a}})}\BibitemShut {NoStop}%
\bibitem [{\citenamefont {Geraedts}\ and\ \citenamefont
  {Motrunich}(2012{\natexlab{b}})}]{short_range3}%
  \BibitemOpen
  \bibfield  {author} {\bibinfo {author} {\bibfnamefont {S.~D.}\ \bibnamefont
  {Geraedts}}\ and\ \bibinfo {author} {\bibfnamefont {O.~I.}\ \bibnamefont
  {Motrunich}},\ }\href {\doibase 10.1103/PhysRevB.86.045106} {\bibfield
  {journal} {\bibinfo  {journal} {Phys. Rev. B}\ }\textbf {\bibinfo {volume}
  {86}},\ \bibinfo {pages} {045106} (\bibinfo {year}
  {2012}{\natexlab{b}})}\BibitemShut {NoStop}%
\bibitem [{\citenamefont {Lee}\ \emph {et~al.}(2016)\citenamefont {Lee},
  \citenamefont {Geraedts},\ and\ \citenamefont {Motrunich}}]{jongyeon}%
  \BibitemOpen
  \bibfield  {author} {\bibinfo {author} {\bibfnamefont {J.~Y.}\ \bibnamefont
  {Lee}}, \bibinfo {author} {\bibfnamefont {S.}~\bibnamefont {Geraedts}}, \
  and\ \bibinfo {author} {\bibfnamefont {O.~I.}\ \bibnamefont {Motrunich}},\
  }\href {\doibase 10.1103/PhysRevB.93.035103} {\bibfield  {journal} {\bibinfo
  {journal} {Phys. Rev. B}\ }\textbf {\bibinfo {volume} {93}},\ \bibinfo
  {pages} {035103} (\bibinfo {year} {2016})}\BibitemShut {NoStop}%
\bibitem [{\citenamefont {Motrunich}\ and\ \citenamefont
  {Vishwanath}(2004)}]{shortlight}%
  \BibitemOpen
  \bibfield  {author} {\bibinfo {author} {\bibfnamefont {O.~I.}\ \bibnamefont
  {Motrunich}}\ and\ \bibinfo {author} {\bibfnamefont {A.}~\bibnamefont
  {Vishwanath}},\ }\href@noop {} {\bibfield  {journal} {\bibinfo  {journal}
  {Phys. Rev. B}\ }\textbf {\bibinfo {volume} {70}},\ \bibinfo {pages} {075104}
  (\bibinfo {year} {2004})}\BibitemShut {NoStop}%
\bibitem [{\citenamefont {Senthil}\ \emph
  {et~al.}(2004{\natexlab{a}})\citenamefont {Senthil}, \citenamefont
  {Vishwanath}, \citenamefont {Balents}, \citenamefont {Sachdev},\ and\
  \citenamefont {Fisher}}]{deccp_science}%
  \BibitemOpen
  \bibfield  {author} {\bibinfo {author} {\bibfnamefont {T.}~\bibnamefont
  {Senthil}}, \bibinfo {author} {\bibfnamefont {A.}~\bibnamefont {Vishwanath}},
  \bibinfo {author} {\bibfnamefont {L.}~\bibnamefont {Balents}}, \bibinfo
  {author} {\bibfnamefont {S.}~\bibnamefont {Sachdev}}, \ and\ \bibinfo
  {author} {\bibfnamefont {M.~P.~A.}\ \bibnamefont {Fisher}},\ }\href@noop {}
  {\bibfield  {journal} {\bibinfo  {journal} {Science}\ }\textbf {\bibinfo
  {volume} {303}},\ \bibinfo {pages} {1490} (\bibinfo {year}
  {2004}{\natexlab{a}})}\BibitemShut {NoStop}%
\bibitem [{\citenamefont {Senthil}\ \emph
  {et~al.}(2004{\natexlab{b}})\citenamefont {Senthil}, \citenamefont {Balents},
  \citenamefont {Sachdev}, \citenamefont {Vishwanath},\ and\ \citenamefont
  {Fisher}}]{deccp_prb}%
  \BibitemOpen
  \bibfield  {author} {\bibinfo {author} {\bibfnamefont {T.}~\bibnamefont
  {Senthil}}, \bibinfo {author} {\bibfnamefont {L.}~\bibnamefont {Balents}},
  \bibinfo {author} {\bibfnamefont {S.}~\bibnamefont {Sachdev}}, \bibinfo
  {author} {\bibfnamefont {A.}~\bibnamefont {Vishwanath}}, \ and\ \bibinfo
  {author} {\bibfnamefont {M.~P.~A.}\ \bibnamefont {Fisher}},\ }\href@noop {}
  {\bibfield  {journal} {\bibinfo  {journal} {Phys. Rev. B}\ }\textbf {\bibinfo
  {volume} {70}},\ \bibinfo {pages} {144407} (\bibinfo {year}
  {2004}{\natexlab{b}})}\BibitemShut {NoStop}%
\bibitem [{\citenamefont {Son}(2015)}]{Son}%
  \BibitemOpen
  \bibfield  {author} {\bibinfo {author} {\bibfnamefont {D.~T.}\ \bibnamefont
  {Son}},\ }\href {\doibase 10.1103/PhysRevX.5.031027} {\bibfield  {journal}
  {\bibinfo  {journal} {Phys. Rev. X}\ }\textbf {\bibinfo {volume} {5}},\
  \bibinfo {pages} {031027} (\bibinfo {year} {2015})}\BibitemShut {NoStop}%
\bibitem [{\citenamefont {Wang}\ and\ \citenamefont
  {Senthil}(2015)}]{WangSenthil2015}%
  \BibitemOpen
  \bibfield  {author} {\bibinfo {author} {\bibfnamefont {C.}~\bibnamefont
  {Wang}}\ and\ \bibinfo {author} {\bibfnamefont {T.}~\bibnamefont {Senthil}},\
  }\href {\doibase 10.1103/PhysRevX.5.041031} {\bibfield  {journal} {\bibinfo
  {journal} {Phys. Rev. X}\ }\textbf {\bibinfo {volume} {5}},\ \bibinfo {pages}
  {041031} (\bibinfo {year} {2015})}\BibitemShut {NoStop}%
\bibitem [{\citenamefont {Metlitski}\ and\ \citenamefont
  {Vishwanath}(2016)}]{MetlitskiVishwanath2015}%
  \BibitemOpen
  \bibfield  {author} {\bibinfo {author} {\bibfnamefont {M.~A.}\ \bibnamefont
  {Metlitski}}\ and\ \bibinfo {author} {\bibfnamefont {A.}~\bibnamefont
  {Vishwanath}},\ }\href {\doibase 10.1103/PhysRevB.93.245151} {\bibfield
  {journal} {\bibinfo  {journal} {Phys. Rev. B}\ }\textbf {\bibinfo {volume}
  {93}},\ \bibinfo {pages} {245151} (\bibinfo {year} {2016})}\BibitemShut
  {NoStop}%
\bibitem [{\citenamefont {Mross}\ \emph {et~al.}(2016)\citenamefont {Mross},
  \citenamefont {Alicea},\ and\ \citenamefont
  {Motrunich}}]{Mross2016_diracduality}%
  \BibitemOpen
  \bibfield  {author} {\bibinfo {author} {\bibfnamefont {D.~F.}\ \bibnamefont
  {Mross}}, \bibinfo {author} {\bibfnamefont {J.}~\bibnamefont {Alicea}}, \
  and\ \bibinfo {author} {\bibfnamefont {O.~I.}\ \bibnamefont {Motrunich}},\
  }\href {\doibase 10.1103/PhysRevLett.117.016802} {\bibfield  {journal}
  {\bibinfo  {journal} {Phys. Rev. Lett.}\ }\textbf {\bibinfo {volume} {117}},\
  \bibinfo {pages} {016802} (\bibinfo {year} {2016})}\BibitemShut {NoStop}%
\bibitem [{\citenamefont {Seiberg}\ \emph {et~al.}(2016)\citenamefont
  {Seiberg}, \citenamefont {Senthil}, \citenamefont {Wang},\ and\ \citenamefont
  {Witten}}]{SeibergSenthilWangWitten2016}%
  \BibitemOpen
  \bibfield  {author} {\bibinfo {author} {\bibfnamefont {N.}~\bibnamefont
  {Seiberg}}, \bibinfo {author} {\bibfnamefont {T.}~\bibnamefont {Senthil}},
  \bibinfo {author} {\bibfnamefont {C.}~\bibnamefont {Wang}}, \ and\ \bibinfo
  {author} {\bibfnamefont {E.}~\bibnamefont {Witten}},\ }\href {\doibase
  http://dx.doi.org/10.1016/j.aop.2016.08.007} {\bibfield  {journal} {\bibinfo
  {journal} {Annals of Physics}\ }\textbf {\bibinfo {volume} {374}},\ \bibinfo
  {pages} {395 } (\bibinfo {year} {2016})}\BibitemShut {NoStop}%
\bibitem [{\citenamefont {Karch}\ and\ \citenamefont
  {Tong}(2016)}]{KarchTong2016}%
  \BibitemOpen
  \bibfield  {author} {\bibinfo {author} {\bibfnamefont {A.}~\bibnamefont
  {Karch}}\ and\ \bibinfo {author} {\bibfnamefont {D.}~\bibnamefont {Tong}},\
  }\href {\doibase 10.1103/PhysRevX.6.031043} {\bibfield  {journal} {\bibinfo
  {journal} {Phys. Rev. X}\ }\textbf {\bibinfo {volume} {6}},\ \bibinfo {pages}
  {031043} (\bibinfo {year} {2016})}\BibitemShut {NoStop}%
\bibitem [{\citenamefont {{Wang}}\ \emph
  {et~al.}(2017{\natexlab{a}})\citenamefont {{Wang}}, \citenamefont {{Nahum}},
  \citenamefont {{Metlitski}}, \citenamefont {{Xu}},\ and\ \citenamefont
  {{Senthil}}}]{WangNahum2017_dccdual}%
  \BibitemOpen
  \bibfield  {author} {\bibinfo {author} {\bibfnamefont {C.}~\bibnamefont
  {{Wang}}}, \bibinfo {author} {\bibfnamefont {A.}~\bibnamefont {{Nahum}}},
  \bibinfo {author} {\bibfnamefont {M.~A.}\ \bibnamefont {{Metlitski}}},
  \bibinfo {author} {\bibfnamefont {C.}~\bibnamefont {{Xu}}}, \ and\ \bibinfo
  {author} {\bibfnamefont {T.}~\bibnamefont {{Senthil}}},\ }\href@noop {}
  {\bibfield  {journal} {\bibinfo  {journal} {ArXiv e-prints}\ } (\bibinfo
  {year} {2017}{\natexlab{a}})},\ \Eprint {http://arxiv.org/abs/1703.02426}
  {arXiv:1703.02426 [cond-mat.str-el]} \BibitemShut {NoStop}%
\bibitem [{\citenamefont {Geraedts}\ \emph {et~al.}(2016)\citenamefont
  {Geraedts}, \citenamefont {Zaletel}, \citenamefont {Mong}, \citenamefont
  {Metlitski}, \citenamefont {Vishwanath},\ and\ \citenamefont
  {Motrunich}}]{Geraedts2016_science}%
  \BibitemOpen
  \bibfield  {author} {\bibinfo {author} {\bibfnamefont {S.~D.}\ \bibnamefont
  {Geraedts}}, \bibinfo {author} {\bibfnamefont {M.~P.}\ \bibnamefont
  {Zaletel}}, \bibinfo {author} {\bibfnamefont {R.~S.~K.}\ \bibnamefont
  {Mong}}, \bibinfo {author} {\bibfnamefont {M.~A.}\ \bibnamefont {Metlitski}},
  \bibinfo {author} {\bibfnamefont {A.}~\bibnamefont {Vishwanath}}, \ and\
  \bibinfo {author} {\bibfnamefont {O.~I.}\ \bibnamefont {Motrunich}},\ }\href
  {\doibase 10.1126/science.aad4302} {\bibfield  {journal} {\bibinfo  {journal}
  {Science}\ }\textbf {\bibinfo {volume} {352}},\ \bibinfo {pages} {197}
  (\bibinfo {year} {2016})}\BibitemShut {NoStop}%
\bibitem [{\citenamefont {Wang}\ and\ \citenamefont
  {Senthil}(2016)}]{WangSenthil2016_CFLs}%
  \BibitemOpen
  \bibfield  {author} {\bibinfo {author} {\bibfnamefont {C.}~\bibnamefont
  {Wang}}\ and\ \bibinfo {author} {\bibfnamefont {T.}~\bibnamefont {Senthil}},\
  }\href {\doibase 10.1103/PhysRevB.94.245107} {\bibfield  {journal} {\bibinfo
  {journal} {Phys. Rev. B}\ }\textbf {\bibinfo {volume} {94}},\ \bibinfo
  {pages} {245107} (\bibinfo {year} {2016})}\BibitemShut {NoStop}%
\bibitem [{\citenamefont {Senthil}\ and\ \citenamefont
  {Fisher}(2006)}]{Senthil2006_theta}%
  \BibitemOpen
  \bibfield  {author} {\bibinfo {author} {\bibfnamefont {T.}~\bibnamefont
  {Senthil}}\ and\ \bibinfo {author} {\bibfnamefont {M.~P.~A.}\ \bibnamefont
  {Fisher}},\ }\href@noop {} {\bibfield  {journal} {\bibinfo  {journal} {Phys.
  Rev. B}\ }\textbf {\bibinfo {volume} {74}},\ \bibinfo {pages} {064405}
  (\bibinfo {year} {2006})}\BibitemShut {NoStop}%
\bibitem [{\citenamefont {{Wang}}\ \emph
  {et~al.}(2017{\natexlab{b}})\citenamefont {{Wang}}, \citenamefont {{Nahum}},
  \citenamefont {{Metlitski}}, \citenamefont {{Xu}},\ and\ \citenamefont
  {{Senthil}}}]{WangNahum2017}%
  \BibitemOpen
  \bibfield  {author} {\bibinfo {author} {\bibfnamefont {C.}~\bibnamefont
  {{Wang}}}, \bibinfo {author} {\bibfnamefont {A.}~\bibnamefont {{Nahum}}},
  \bibinfo {author} {\bibfnamefont {M.~A.}\ \bibnamefont {{Metlitski}}},
  \bibinfo {author} {\bibfnamefont {C.}~\bibnamefont {{Xu}}}, \ and\ \bibinfo
  {author} {\bibfnamefont {T.}~\bibnamefont {{Senthil}}},\ }\href@noop {}
  {\bibfield  {journal} {\bibinfo  {journal} {ArXiv e-prints}\ } (\bibinfo
  {year} {2017}{\natexlab{b}})},\ \Eprint {http://arxiv.org/abs/1703.02426}
  {arXiv:1703.02426 [cond-mat.str-el]} \BibitemShut {NoStop}%
\bibitem [{\citenamefont {{Mross}}\ \emph {et~al.}(2017)\citenamefont
  {{Mross}}, \citenamefont {{Alicea}},\ and\ \citenamefont
  {{Motrunich}}}]{Mross2017}%
  \BibitemOpen
  \bibfield  {author} {\bibinfo {author} {\bibfnamefont {D.~F.}\ \bibnamefont
  {{Mross}}}, \bibinfo {author} {\bibfnamefont {J.}~\bibnamefont {{Alicea}}}, \
  and\ \bibinfo {author} {\bibfnamefont {O.~I.}\ \bibnamefont {{Motrunich}}},\
  }\href@noop {} {\bibfield  {journal} {\bibinfo  {journal} {ArXiv e-prints}\ }
  (\bibinfo {year} {2017})},\ \Eprint {http://arxiv.org/abs/1705.01106}
  {arXiv:1705.01106 [cond-mat.str-el]} \BibitemShut {NoStop}%
\bibitem [{Note1()}]{Note1}%
  \BibitemOpen
  \bibinfo {note} {The transition from the trivial insulator to phase where
  both species are superfluid at $\eta = 2/5$ was not the main focus of
  Ref.~\protect \rev@citealpnum {jongyeon} and was not determined accurately;
  in particular, it was determined by looking at specific heat as a function of
  $\lambda _1$ with $\lambda _2$ fixed. In this work we studied convenient
  ``superfluid stiffness'' (i.e., current-current correlation in convenient
  simulation variables) as a function of $\lambda _1 = \lambda _2$ (similar to
  Fig.~4 of Ref.~\protect \rev@citealpnum {short_range3}), obtaining a more
  accurate location of the transition which is plotted in Fig.~\ref
  {fig:fixedeta}. We obtain a transition at $\lambda _c \approx 0.27$, while in
  Ref.~\protect \rev@citealpnum {jongyeon} we reported a value of $\lambda _c
  \approx 0.23$.}\BibitemShut {Stop}%
\bibitem [{Note2()}]{Note2}%
  \BibitemOpen
  \bibinfo {note} {In particular, one can take Eq.~(\ref {action}) and use
  interactions which are $\propto 1/k$ combined with non-local $w(k) = \eta
  '/(2\pi |f_k|^2)$. Doing this gives the model an additional property that the
  interactions have the same form in terms of both $\protect \mathaccentV
  {vec}17E{\protect \mathcal {J}}$ and $\protect \mathaccentV {vec}17E{\protect
  \mathcal {Q}}$ variables. The combination of dualizing $\protect \mathaccentV
  {vec}17E{\protect \mathcal {J}}_{1,2}$ to $\protect \mathaccentV
  {vec}17E{\protect \mathcal {Q}}_{1,2}$ and shifting $\eta '$ by $1$ (another
  property of the model derived from the fact that in the lattice loop model
  shifting the statistical angle by $2\pi $ does not change the contribution to
  the partition sum) generates the entires modular group, and this allows the
  phase diagram to be determined analytically.\cite {Gen2Loops}}\BibitemShut
  {NoStop}%
\bibitem [{\citenamefont {Fradkin}\ and\ \citenamefont
  {Kivelson}(1996)}]{Fradkin_SL2Z}%
  \BibitemOpen
  \bibfield  {author} {\bibinfo {author} {\bibfnamefont {E.}~\bibnamefont
  {Fradkin}}\ and\ \bibinfo {author} {\bibfnamefont {S.}~\bibnamefont
  {Kivelson}},\ }\href@noop {} {\bibfield  {journal} {\bibinfo  {journal}
  {Nucl. Phys. B}\ }\textbf {\bibinfo {volume} {474}},\ \bibinfo {pages} {543}
  (\bibinfo {year} {1996})}\BibitemShut {NoStop}%
\bibitem [{\citenamefont {Geraedts}\ and\ \citenamefont
  {Motrunich}(2012{\natexlab{c}})}]{Gen2Loops}%
  \BibitemOpen
  \bibfield  {author} {\bibinfo {author} {\bibfnamefont {S.~D.}\ \bibnamefont
  {Geraedts}}\ and\ \bibinfo {author} {\bibfnamefont {O.~I.}\ \bibnamefont
  {Motrunich}},\ }\href {\doibase 10.1103/PhysRevB.86.245121} {\bibfield
  {journal} {\bibinfo  {journal} {Phys. Rev. B}\ }\textbf {\bibinfo {volume}
  {86}},\ \bibinfo {pages} {245121} (\bibinfo {year}
  {2012}{\natexlab{c}})}\BibitemShut {NoStop}%
\bibitem [{\citenamefont {Kuklov}\ \emph {et~al.}(2006)\citenamefont {Kuklov},
  \citenamefont {Prokof'ev}, \citenamefont {Svistunov},\ and\ \citenamefont
  {Troyer}}]{Kuklov06}%
  \BibitemOpen
  \bibfield  {author} {\bibinfo {author} {\bibfnamefont {A.~B.}\ \bibnamefont
  {Kuklov}}, \bibinfo {author} {\bibfnamefont {N.~V.}\ \bibnamefont
  {Prokof'ev}}, \bibinfo {author} {\bibfnamefont {B.~V.}\ \bibnamefont
  {Svistunov}}, \ and\ \bibinfo {author} {\bibfnamefont {M.}~\bibnamefont
  {Troyer}},\ }\href@noop {} {\bibfield  {journal} {\bibinfo  {journal} {Ann.
  Phys. (N.Y.)}\ }\textbf {\bibinfo {volume} {321}},\ \bibinfo {pages} {1602}
  (\bibinfo {year} {2006})}\BibitemShut {NoStop}%
\bibitem [{\citenamefont {Smiseth}\ \emph {et~al.}(2005)\citenamefont
  {Smiseth}, \citenamefont {Smorgrav}, \citenamefont {Babaev},\ and\
  \citenamefont {Sudbo}}]{Smiseth05}%
  \BibitemOpen
  \bibfield  {author} {\bibinfo {author} {\bibfnamefont {J.}~\bibnamefont
  {Smiseth}}, \bibinfo {author} {\bibfnamefont {E.}~\bibnamefont {Smorgrav}},
  \bibinfo {author} {\bibfnamefont {E.}~\bibnamefont {Babaev}}, \ and\ \bibinfo
  {author} {\bibfnamefont {A.}~\bibnamefont {Sudbo}},\ }\href@noop {}
  {\bibfield  {journal} {\bibinfo  {journal} {Phys. Rev. B}\ }\textbf {\bibinfo
  {volume} {71}},\ \bibinfo {pages} {214509} (\bibinfo {year}
  {2005})}\BibitemShut {NoStop}%
\bibitem [{\citenamefont {Kragset}\ \emph {et~al.}(2006)\citenamefont
  {Kragset}, \citenamefont {Smorgrav}, \citenamefont {Hove}, \citenamefont
  {Nogueira},\ and\ \citenamefont {Sudbo}}]{Kragset06}%
  \BibitemOpen
  \bibfield  {author} {\bibinfo {author} {\bibfnamefont {S.}~\bibnamefont
  {Kragset}}, \bibinfo {author} {\bibfnamefont {E.}~\bibnamefont {Smorgrav}},
  \bibinfo {author} {\bibfnamefont {J.}~\bibnamefont {Hove}}, \bibinfo {author}
  {\bibfnamefont {F.~S.}\ \bibnamefont {Nogueira}}, \ and\ \bibinfo {author}
  {\bibfnamefont {A.}~\bibnamefont {Sudbo}},\ }\href@noop {} {\bibfield
  {journal} {\bibinfo  {journal} {Phys. Rev. Lett.}\ }\textbf {\bibinfo
  {volume} {97}},\ \bibinfo {pages} {247201} (\bibinfo {year}
  {2006})}\BibitemShut {NoStop}%
\bibitem [{\citenamefont {Herland}\ \emph {et~al.}(2010)\citenamefont
  {Herland}, \citenamefont {Babaev},\ and\ \citenamefont
  {Sudb\o{}}}]{Herland2010}%
  \BibitemOpen
  \bibfield  {author} {\bibinfo {author} {\bibfnamefont {E.~V.}\ \bibnamefont
  {Herland}}, \bibinfo {author} {\bibfnamefont {E.}~\bibnamefont {Babaev}}, \
  and\ \bibinfo {author} {\bibfnamefont {A.}~\bibnamefont {Sudb\o{}}},\ }\href
  {\doibase 10.1103/PhysRevB.82.134511} {\bibfield  {journal} {\bibinfo
  {journal} {Phys. Rev. B}\ }\textbf {\bibinfo {volume} {82}},\ \bibinfo
  {pages} {134511} (\bibinfo {year} {2010})}\BibitemShut {NoStop}%
\bibitem [{\citenamefont {Kitaev}(2003)}]{Kitaev2003}%
  \BibitemOpen
  \bibfield  {author} {\bibinfo {author} {\bibfnamefont {A.}~\bibnamefont
  {Kitaev}},\ }\href@noop {} {\bibfield  {journal} {\bibinfo  {journal} {Annals
  of Physics}\ }\textbf {\bibinfo {volume} {303}},\ \bibinfo {pages} {2}
  (\bibinfo {year} {2003})}\BibitemShut {NoStop}%
\end{thebibliography}%

\end{document}